\documentclass{aa}  

\usepackage{graphicx}
\usepackage{txfonts}
\usepackage{natbib}

\usepackage{aalongtable,lscape}
\usepackage{longtable}

\newcommand{\teff}{T$_{\rm eff}$ }
\newcommand{\tsin}{T$_{\rm eff}$}

\begin{document}

\title{Chemical similarities between Galactic bulge and local thick disk red giants: 
O, Na, Mg, Al, Si, Ca and Ti}

\titlerunning{Chemical similarities between the Galactic bulge and thick disk}
\authorrunning{Alves-Brito et al.}

\author{
A.~Alves-Brito\inst{1,2} \and
J.~Mel\'{e}ndez\inst{3} \and
M.~Asplund\inst{4} \and
I.~Ram\'{\i}rez \inst{4} \and
D.~Yong\inst{5}
}


\institute{
Universidade de S\~{a}o Paulo, IAG, Rua do Mat\~{a}o 1226, 
Cidade Universit\'{a}ria, S\~{a}o Paulo 05508-900, Brazil;
\email{abrito@astro.iag.usp.br} \and
Centre for Astrophysics and Supercomputing, Swinburne University of
Technology, Hawthorn, Victoria 3122, Australia\and
Centro de Astrof\'{\i}sica da Universidade do Porto, 
Rua das Estrelas, 4150-762 Porto, Portugal; \email{jorge@astro.up.pt} 
\and
Max Planck Institut f\"ur Astrophysik,
Postfach 1317, 85741 Garching, Germany
\and
Research School of Astronomy and Astrophysics,
The Australian National University, Cotter Road, Weston, ACT 2611, Australia
}

\date{Received: ; accepted: }

\abstract
{The formation and evolution of the Galactic bulge and its relationship
with the other Galactic populations is still poorly understood.
}
{To establish the chemical differences and similarities between 
the bulge and other stellar populations, we performed an 
elemental abundance analysis of $\alpha$- (O, Mg, Si, Ca, and Ti) and
Z-odd (Na and Al) elements of red giant stars in the bulge as well as of
local thin disk, thick disk and halo giants.
}
{We use high-resolution 
optical spectra of 25 bulge giants in Baade's window and 55 comparison
giants (4 halo, 29 thin disk and 22 thick disk giants) in the solar neighborhood.
All stars have similar stellar parameters but cover a broad range in metallicity
($-1.5 <$ [Fe/H] $< +0.5$). 
A standard 1D local thermodynamic equilibrium analysis using
both Kurucz and MARCS models yielded the abundances of 
O, Na, Mg, Al, Si, Ca, Ti and Fe.
Our homogeneous and differential analysis of the Galactic stellar populations 
ensured that systematic errors were minimized.}
{We confirm the well-established differences for [$\alpha$/Fe] at a given
metallicity between the local thin and thick disks. 
For all the elements investigated, we find no chemical distinction 
between the bulge and the local thick disk, in agreement with our previous study
of C, N and O but in contrast to other groups relying on literature values for nearby disk dwarf stars.
For  $-1.5 <$ [Fe/H] $< -0.3$ 
exactly the same trend is followed by both the bulge and thick disk stars, with
a star-to-star scatter of only 0.03 dex. 
Furthermore, both populations share the location of the knee in the 
[$\alpha$/Fe] vs [Fe/H] diagram. It still remains to be confirmed
that the local thick disk extends to super-solar metallicities as is the case for the bulge. 
These are the most stringent constraints to date on 
the chemical similarity of these stellar populations.
}
{Our findings suggest that the bulge and local thick disk stars
experienced similar formation timescales, star formation rates and
initial mass functions, confirming thus the main outcomes
of our previous homogeneous analysis of  [O/Fe] from infrared spectra 
for nearly the same sample. 
The identical $\alpha$-enhancements of thick disk and bulge stars may reflect a 
rapid chemical evolution taking place before the 
bulge and thick disk structures we see today were formed, or it may reflect
Galactic orbital
migration of inner disk/bulge stars resulting in stars in the solar neighborhood
with thick-disk kinematics.
}


\maketitle
%

\section{Introduction}

The Galactic bulge is the least understood stellar population in the Milky Way, as even 
its classification (classical or pseudo-bulge; Kormendy \& Kennicutt 2004) seems unclear. 
The Galactic bulge has signatures of an old 
(Ortolani et al. 1995; Zoccali et al. 2003) classical bulge formed rapidly during 
intensive star formation as reflected in the enhancement of 
$\alpha$-elements (e.g. McWilliam \& Rich 1994; Cunha \& Smith 2006; Zoccali et al. 2006; 
Lecureur et al. 2007; Fulbright et al. 2007; Mel\'endez et al. 2008; Ryde et al. 2009a,b).
On the other hand its boxy shape is consistent with a pseudo-bulge 
indicative of formation by secular evolution through dynamical instability of 
an already established inner disk. 

Recently, Elmegreen et al. (2008) have shown that bulges formed by coalescence of 
giant clumps can have properties of both classical and pseudo-bulges, because secular 
evolution can take place in a very short timescale ($<$ 1 Gyr). They suggest that our 
Galactic bulge (and many $z \sim 2$ early disk galaxies; Genzel et al. 2008) formed this way, 
and that the bulge and thick disk may have formed at the same time.
Thus, the nature of our Galactic bulge can be unveiled 
by detailed chemical composition analysis and by careful comparisons with the thick disk. 

Although all recent works agree in enhancements of the $\alpha$-elements relative to 
solar abundances in bulge 
field K giants, the level of enhancement is currently under debate. Based on a 
comparison of bulge giant stars with thick disk dwarf stars, Zoccali et al. (2006), 
Lecureur et al. (2007) and Fulbright et al. (2007) suggested that the bulge and the 
thick disk have different chemical composition patterns, and that the $\alpha$-elements 
are overabundant in the bulge compared with the thick disk. Therefore, they
argued for a shorter formation timescale and higher star formation rate for the Galactic bulge 
than that for the thick disk.
Ballero et al. (2007) also concluded that the initial mass functions 
must have been different between the two populations 
based on both the high [Mg/Fe] and metallicity distribution of the bulge
(see also Cescutti et al. 2009).
Nevertheless, those comparisons should be taken with care as systematic errors may 
be present due to the very different stellar parameters, model atmospheres, and 
NLTE effects of dwarf and giant stars. Indeed, in our consistent analysis of high 
resolution infrared spectra of both bulge and thick disk giants with similar stellar parameters
(Mel\'endez et al. 2008), we have shown that the bulge 
is in fact chemically very similar to the thick disk in [C/Fe], [N/Fe] and [O/Fe].
Here, we extend this work to other $\alpha$-elements (Mg, Si, Ti, Ca), and show that 
all the $\alpha$-elements in bulge and local thick disk giants have essentially 
identical chemical abundance patterns.

\begin{figure}
\centering
\includegraphics[width=8cm]{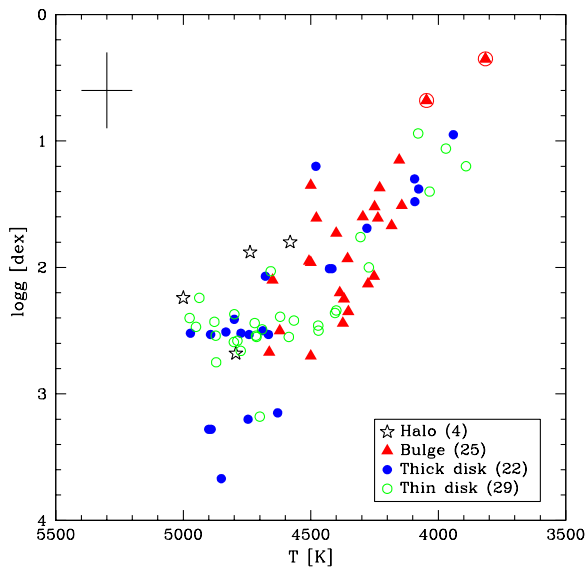}
\caption{H-R diagram showing our program stars. 
The symbols are described in the plot. The two
most luminous stars (filled triangles enclosed by circles) 
are bulge giants which show abundance anomalies like O-deficiency and Na-enhancement
similar to those observed in some globular cluster stars.
A typical error bar in $T_{\rm eff}$ and $\log g$ is shown.}
\label{f:cmd}
 \end{figure}

\section{Observations}
\label{s:observations}
The sample consists of 80 cool giant stars (Fig. \ref{f:cmd}) with 
effective temperatures $3800 \le T_{\rm eff} \le 5000$\,K,
surface gravities $0.5 \le \log g \le 3.5$, 
and metallicities $-1.5 <$[Fe/H]$<+0.5$. Similar number of thin disk (29), thick disk (22) 
and bulge (25) giants were selected, and a few (4) metal-rich halo giants were also included. 

All of our bulge giants are located in Baade's window and are taken 
from Fulbright et al. (2006), who have cleaned the sample from nonbulge giants.
For these bulge stars, we have already published an abundance analysis of C, N, O and Fe 
based on IR spectra (Mel\'endez et al. 2008). For the present study we make use of 
the equivalent widths measured in optical spectra
using the HIRES spectrograph (at R = 45,000 or 67,000) on the Keck-I 10\,m telescope
by  \citet{2006ApJ...636..821F,  2007ApJ...661.1152F}. 

To enable a proper comparison we have compiled a sample of thin disk, thick disk
and halo stars for which we have obtained our own optical spectra. 
The assignment of population membership was based on 
UVW velocities \citep{2004A&A...415..155B, 2006MNRAS.367.1329R}. The sample selection 
was based on evaluating population membership in more than 1500 giant stars 
from the literature, in particular an updated version of the 
Cayrel de Strobel (2001) catalog (see Ram\'{\i}rez \& Mel\'endez 2005a), 
the analysis of $\sim$180 clump giants by Mishenina et al. (2006), the study of $\sim$300 
nearby giants by Luck \& Heiter (2007), the survey of $\sim$380 giants by 
Hekker \& Mel\'endez (2007), and the analysis of $\sim$320 giants by 
Takeda et al. (2008). Furthermore, the UVES library of stellar spectra
(Bagnulo et al. 2003) was searched for suitable disk and halo giants.

Our analysis of thin disk, thick disk and halo stars is based mostly on 
high-resolution (R= 65,000) optical spectra taken 
in April 2007 with the MIKE spectrograph (Bernstein et al. 2003) on the 
Clay 6.5\,m Magellan telescope, and complemented with observations using 
the 2dcoud\'e  spectrograph (Tull et al. 1995, R = 60,000) on the 
2.7\,m Harlan J. Smith telescope at McDonald Observatory, the upgraded HIRES 
spectrograph (Vogt et al. 1994, R = 100,000) on the Keck I 10\,m telescope, 
the UVES library\footnote{http://www.sc.eso.org/santiago/uvespop/} (Bagnulo et al. 2003, R = 80,000), and the 
ELODIE archive\footnote{http://atlas.obs-hp.fr/elodie/} (Moultaka et al. 2004, R = 42,000). 
The magnitudes, population membership and instrumentation used for the
disk/halo sample are shown in Table \ref{t:stars}.

The data were reduced with IRAF employing standard procedures: correction 
for bias, flat field, cosmic rays and background light, 
then optimal extraction of the spectra (using a bright star to trace the orders), 
wavelength calibration, barycentric and Doppler correction, and continuum 
normalization. In some cases, as described below, a variation to the reduction procedure was 
necessary. The tilt of the lines in the MIKE data is severe 
and varies across the CCD \citep[e.g.][]{2006ApJ...639..918Y}, therefore it must be 
carefully corrected to avoid degradation of the spectral resolution. The tilt was 
corrected using 
MTOOLS\footnote{http://www.lco.cl/telescopes-information/magellan/instruments-1/mike/IRAF\_tools/iraf-mtools-package/}, 
specifically developed by J. Baldwin to account for the tilted slits in MIKE spectra. 
On the other hand, our HIRES spectra were extracted using a new version of 
MAKEE\footnote{http://www2.keck.hawaii.edu/inst/hires/hires.html}, 
an optimal extraction package developed by T. Barlow specifically for data reduction 
of the improved HIRES spectrograph. MAKEE also performs an automatic wavelength 
calibration cross-correlating the extracted ThAr spectra with a database of 
wavelength calibration solutions. Both the UVES and ELODIE archive data were 
already extracted and wavelength calibrated. The extracted spectra
were shifted to the rest frame and continuum normalized using IRAF. 
The signal-to-noise ratio (S/N) per pixel of the reduced spectra ranges 
from $S/N \sim 45-100$ for the bulge giants (Fulbright et al. 2006),
whereas for the disk and halo giants the S/N is typically $\sim$200 per pixel,
ranging from $\sim$150 (2dCoude/McDonald) to $\sim$200 (MIKE/Magellan, ELODIE/OHP) to
$\sim$250 (HIRES/Keck, UVES/VLT), as estimated from relatively line-free regions of the spectra.
\begin{table}
\begin{flushleft}
\caption{Program stars data}
\label{t:stars}      
\centering          
\begin{tabular}{lcllll}     
\noalign{\smallskip}
\hline\hline    
\noalign{\smallskip}
\noalign{\vskip 0.1cm} 
Star & V [mag]  & P$^{*}$ [\%] & Instrument \\        
(1) & (2)  & (3)  & (4)   \\                    
\noalign{\vskip 0.1cm}
\noalign{\hrule\vskip 0.1cm}
\noalign{\vskip 0.1cm}    
\multicolumn{4}{c}{\hbox{\bf Halo}} \\
\noalign{\vskip 0.1cm}
\noalign{\hrule\vskip 0.1cm}
\noalign{\vskip 0.1cm} 
HD041667 & 8.533  & 00:01:99  & MIKE/Magellan \\ 
HD078050 & 7.676  & 00:00:100 & ELODIE/OHP        \\
HD114095 & 8.353  & 00:29:71  & MIKE/Magellan       \\
HD210295 & 9.566  & 00:00:100 & HIRES/Keck	       \\
\noalign{\smallskip}
\noalign{\vskip 0.1cm}
\noalign{\hrule\vskip 0.1cm}
\noalign{\vskip 0.1cm}
\multicolumn{4}{c}{\hbox{\bf Thick~Disk}}\\
\noalign{\vskip 0.1cm}
\noalign{\hrule\vskip 0.1cm}
\noalign{\vskip 0.1cm} 
HD023940   &   5.541    &  03:96:01   &   2dcoude/McDonald  \\ 
HD032440   &   5.459    &  08:91:01   &   MIKE/Magellan\\
HD037763   &   5.178    &  15:84:01   &   MIKE/Magellan \\
HD040409   &   4.645    &  25:74:01   &   MIKE/Magellan \\
HD077236   &   7.499    &  00:58:42   &   MIKE/Magellan \\
HD077729   &   7.630    &  25:74:01   &   MIKE/Magellan  \\
HD080811   &   8.35     &  00:97:03   &   MIKE/Magellan  \\
HD083212   &   8.335    &  00:94:06   &   2dcoude/McDonald  \\
HD099978   &   8.653    &  00:99:01   &   MIKE/Magellan\\
HD107328   &   4.967    &  43:57:01   &   2dcoude/McDonald  \\
HD107773   &   6.355    &  20:78:02   &   MIKE/Magellan  \\
HD119971   &   5.454    &  14:85:01   &   MIKE/Magellan  \\
HD124897   &  -0.049    &  13:85:01   &   MIKE/Magellan  \\
HD127243   &   5.590    &  00:96:04   &   ELODIE/OHP  \\
HD130952   &   4.943    &  01:98:01   &   MIKE/Magellan  \\
HD136014   &   6.195    &  24:75:01   &   MIKE/Magellan  \\
HD145148   &   5.954    &  11:88:01   &   MIKE/Magellan  \\
HD148451   &   6.564    &  00:64:36   &   UVES/VLT  \\
HD180928   &   6.088    &  00:74:26   &   MIKE/Magellan  \\
HD203344   &   5.570    &  00:97:02   &   ELODIE/OHP  \\
HD219615   &   3.694    &  29:70:01   &   ELODIE/OHP  \\
HD221345   &   5.220    &  14:85:01   &   ELODIE/OHP  \\
\noalign{\vskip 0.1cm}						  
\noalign{\hrule\vskip 0.1cm}
\noalign{\vskip 0.1cm}  
\multicolumn{4}{c}{\hbox{\bf Thin~Disk}} \\
\noalign{\vskip 0.1cm}
\noalign{\hrule\vskip 0.1cm}
\noalign{\vskip 0.1cm}  
HD000787  &   5.255  & 98:02:00  &  2dcoude/McDonald	       \\ 
HD003546  &   4.361  & 78:22:00  &  ELODIE/OHP	       \\
HD005268  &   6.163  & 86:14:00  &  HIRES/Keck	       \\
HD029139  &   0.868  & 98:02:00  &  ELODIE/OHP	       \\
HD029503  &   3.861  & 96:04:00  &  2dcoude/McDonald	       \\
HD030608  &   6.362  & 49:51:00  &  2dcoude/McDonald	       \\
HD045415  &   5.543  & 99:01:00  &  UVES/VLT	       \\
HD050778  &   4.065  & 87:13:00  &  2dcoude/McDonald	       \\
HD073017  &   5.673  & 86:14:00  &  ELODIE/OHP	       \\
HD099648  &   4.952  & 99:01:00  &  MIKE/Magellan	       \\
HD100920  &   4.301  & 99:01:00  &  MIKE/Magellan	       \\
HD115478  &   5.333  & 99:01:00  &  MIKE/Magellan	       \\
HD116976  &   4.753  & 99:01:00  &  MIKE/Magellan	       \\
HD117220  &   9.010  & 95:05:00  &  MIKE/Magellan	       \\
HD117818  &   5.205  & 99:01:00  &  MIKE/Magellan	       \\
HD128188  &   10.003 & 98:02:00  &  MIKE/Magellan	       \\
HD132345  &   5.838  & 97:03:00  &  MIKE/Magellan	       \\
HD142948  &   8.024  & 94:06:00  &  MIKE/Magellan	       \\
HD171496  &   8.501  & 98:02:00  &  2dcoude/McDonald	       \\
HD172223  &   6.485  & 91:09:00  &  MIKE/Magellan	       \\
HD174116  &   5.24   & 98:02:00  &  2dcoude/McDonald	       \\
HD175219  &   5.355  & 99:01:00  &  MIKE/Magellan	       \\
HD186378  &   7.21   & 97:03:00  &  2dcoude/McDonald	       \\
HD187195  &   6.022  & 99:01:00  &  MIKE/Magellan	       \\
HD211075  &   8.190  & 99:01:00  &  HIRES/Keck	       \\
HD212320  &   5.938  & 99:01:00  &  UVES/VLT   \\
HD214376  &   5.036  & 99:01:00  &  HIRES/Keck	       \\
HD215030  &   5.92   & 98:02:00  &  ELODIE/OHP	       \\
HD221148  &   6.252  & 93:07:00  &  HIRES/Keck	       \\
\hline                  
\end{tabular}
\begin{minipage}{.88\hsize}
 Notes.--- (*): The membership probabilities of the
thin disk, thick disk and halo giants are given as thin:thick:halo.\\
\end{minipage}			
\end{flushleft}
\end{table}  

\begin{table}
\begin{flushleft}
\caption{Sensitivities in the abundance ratios by employing the Kurucz models
(Castelli et al. 1997). 
The atmospheric parameters and $\alpha$-enhancement were changed by 
$\Delta T_{\rm eff} = \pm 75$\,K, $\Delta \log g = \pm 0.30$\,dex, 
$\Delta v_{\rm t} = \pm 0.20$\,km\,s$^{-1}$, and
$\Delta [\alpha/{\rm Fe}] = \pm 0.10$\,dex. The total internal uncertainties are
given in the last column}

\label{t:error}      
\centering          
\begin{tabular}{lccccccccccc}     
\noalign{\smallskip}
\hline\hline    
\noalign{\smallskip}
\noalign{\vskip 0.1cm} 
Abundance & $\Delta$T$_{\rm eff}$  & $\Delta$log g & $\Delta$v$_{\rm t}$ &
$\Delta$[$\alpha$/Fe] & ($\sum x^{2}$)$^{1/2}$ \\        
   
(1) & (2)  & (3)  & (4) & (5) & (6)   \\ 
\noalign{\vskip 0.1cm}
\noalign{\hrule\vskip 0.1cm}
\noalign{\vskip 0.1cm}    
\multicolumn{6}{c}{\hbox{\bf HD078050}} \\
\noalign{\vskip 0.1cm}
\noalign{\hrule\vskip 0.1cm}
\noalign{\vskip 0.1cm} 
\hbox{[FeI/H]}   &    -0.09   &     0.01  &	 0.07	  &   0.00 &  0.11 \\
\hbox{[FeII/H]}  &     0.02   &    -0.12  &	 0.06	  &  -0.02 &  0.14 \\
\hbox{[O/Fe]}    &    -0.01   &    -0.13  &	 0.00	  &  -0.02 &  0.13 \\
\hbox{[Na/Fe]}   &    -0.06   &     0.01  &	 0.01	  &   0.00 &  0.06 \\
\hbox{[Mg/Fe]}   &    -0.06   &     0.05  &	 0.04	  &   0.00 &  0.09 \\
\hbox{[Al/Fe]}   &    -0.06   &     0.00  &	 0.00	  &   0.00 &  0.06 \\
\hbox{[Si/Fe]}   &    -0.03   &    -0.03  &	 0.01	  &   0.00 &  0.04 \\
\hbox{[Ca/Fe]}   &    -0.08   &     0.02  &	 0.06	  &   0.00 &  0.10 \\
\hbox{[Ti/Fe]}   &    -0.11   &     0.01  &	 0.06	  &   0.01 &  0.13 \\			
\noalign{\vskip 0.1cm}
\noalign{\hrule\vskip 0.1cm}
\noalign{\vskip 0.1cm}    
\multicolumn{6}{c}{\hbox{\bf IV203}} \\
\noalign{\vskip 0.1cm}
\noalign{\hrule\vskip 0.1cm}
\noalign{\vskip 0.1cm} 
\hbox{[FeI/H]}   &     0.01   &    -0.07   &	 0.06	  &  -0.01 &  0.09 \\
\hbox{[FeII/H]}  &     0.18   &    -0.18   &	 0.04	  &  -0.02 &  0.26 \\
\hbox{[O/Fe]}	 &    -0.01   &    -0.11   &	 0.01	  &  -0.02 &  0.11 \\
\hbox{[Na/Fe]}   &    -0.08   &     0.04   &	 0.05	  &   0.01 &  0.10 \\
\hbox{[Mg/Fe]}   &     0.01   &    -0.04   &	 0.03	  &  -0.01 &  0.05 \\
\hbox{[Al/Fe]}   &    -0.06   &     0.02   &	 0.03	  &   0.01 &  0.07 \\
\hbox{[Si/Fe]}   &     0.11   &    -0.09   &	 0.03	  &  -0.01 &  0.14 \\
\hbox{[Ca/Fe]}   &    -0.09   &     0.02   &	 0.09	  &   0.01 &  0.13 \\
\hbox{[Ti/Fe]}   &    -0.14   &     0.00   &	 0.03	  &   0.00 &  0.14 \\
\noalign{\vskip 0.1cm}
\noalign{\hrule\vskip 0.1cm}
\noalign{\vskip 0.1cm}    
\multicolumn{6}{c}{\hbox{\bf HD083212}} \\
\noalign{\vskip 0.1cm}
\noalign{\hrule\vskip 0.1cm}
\noalign{\vskip 0.1cm} 

\hbox{[FeI/H]}   &   -0.09  &	  -0.01    &	0.04	 &   0.00 &  0.09  \\
\hbox{[FeII/H]}  &    0.05  &	  -0.12    &	0.05	 &  -0.02 &  0.14  \\
\hbox{[O/Fe]}    &    0.00  &	  -0.13    &	0.01	 &  -0.02 &  0.13  \\
\hbox{[Na/Fe]}   &   -0.07  &	   0.01    &	0.01	 &   0.00 &  0.07  \\
\hbox{[Mg/Fe]}   &   -0.07  &	   0.05    &	0.05	 &   0.00 &  0.09  \\
\hbox{[Si/Fe]}   &   -0.01  &	  -0.04    &	0.00	 &  -0.01 &  0.04  \\
\hbox{[Ca/Fe]}   &   -0.09  &	   0.02    &	0.05	 &   0.01 &  0.10  \\
\hbox{[Ti/Fe]}   &   -0.16  &	   0.00    &	0.07	 &   0.02 &  0.17  \\
\noalign{\vskip 0.1cm}						  
\noalign{\hrule\vskip 0.1cm}
\noalign{\vskip 0.1cm}    
\multicolumn{6}{c}{\hbox{\bf HD045415}} \\
\noalign{\vskip 0.1cm}
\noalign{\hrule\vskip 0.1cm}
\noalign{\vskip 0.1cm} 
\hbox{[FeI/H]}   &   -0.04   &    -0.02  &	0.09	 &  -0.01 &   0.10  \\
\hbox{[FeII/H]}  &    0.08   &    -0.14  &	0.08	 &  -0.03 &   0.18  \\
\hbox{[O/Fe]}    &    0.00   &    -0.14  &	0.00	 &  -0.03 &   0.14  \\
\hbox{[Na/Fe]}   &   -0.06   &     0.08  &	0.05	 &   0.00 &   0.11  \\
\hbox{[Mg/Fe]}   &   -0.02   &     0.02  &	0.03	 &  -0.01 &   0.04  \\
\hbox{[Al/Fe]}   &   -0.05   &     0.02  &	0.04	 &   0.01 &   0.07  \\
\hbox{[Si/Fe]}   &    0.03   &    -0.05  &	0.03	 &  -0.02 &   0.07  \\
\hbox{[Ca/Fe]}   &   -0.08   &     0.04  &	0.10	 &   0.00 &   0.13  \\
\hbox{[Ti/Fe]}   &   -0.11   &    -0.01  &	0.01	 &   0.00 &   0.11  \\
\hline          	   		  			  
\end{tabular}
\end{flushleft}
\end{table}

\section{Abundance Analysis}
\label{s:analysis}

\begin{figure}
\centering
\includegraphics[width=8cm]{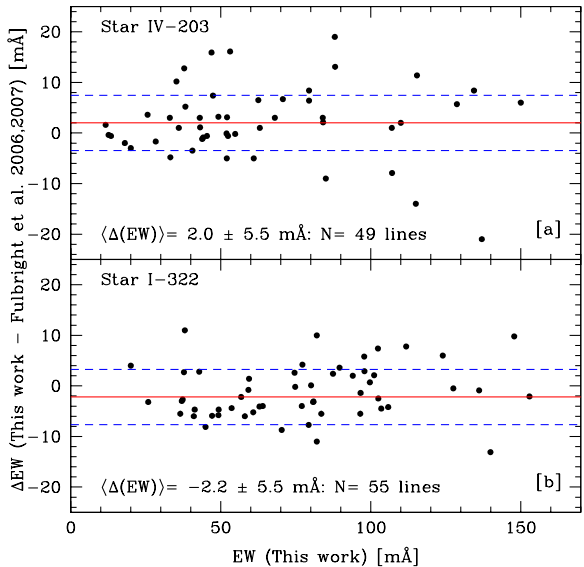}
\caption{Comparison of our equivalent width measurements with lines in
common with Fulbright et al. (2006, 2007) for two bulge giant stars: (a) IV-203 and (b) I-322. 
The median (solid lines) and a robust proxy of standard deviation ($\sigma_{\rm QD}$; 
dashed lines) are also displayed.}
\label{f:iv203}
 \end{figure}

We have homogeneously performed all 
the equivalent width (EW) measurements for the disk and halo sample.
In order to check that the EW measurements of the bulge giants by Fulbright et al. (2006, 2007)
are consistent with our system for the disk and halo giants, 
we have observed one bulge star (IV-203) with the MIKE spectrograph and 
compared the EW measured by us with those obtained by Fulbright et al.
(2006, 2007).
The agreement is satisfactory, with a mean difference (This work -
Fulbright et al. 2006, 2007)
of $-2.0$\,m\AA\ and a 
line-to-line scatter of $\sigma_{\rm QD}\footnote{we use here a robust 
standard deviation based on the quartile deviation QD (=Q3-Q1),
$\sigma_{\rm QD}$ = QD/1.349; see for example Abu-Shawiesh et al. (2009)}
 = 5.5$\,m\AA\ (Fig. \ref{f:iv203}a).
Additionally, A. McWilliam has kindly made available to us 
the HIRES/Keck spectrum of another bulge giant (I-322) for 
comparison purposes. Again, the agreement is good with a difference 
(This work - Fulbright et al. 2006, 2007) of $+2.2$ m$\rm \AA$ and $\sigma_{\rm QD}$ = 5.5 m\AA\ 
(Fig. \ref{f:iv203}b). 
Since most of the employed lines are relatively strong, the typical
impact on abundances from these EW differences is negligible.
Thus, the analysis of 
the faint bulge giants, the bright disk and the halo giants, 
is essentially in the same system.

\begin{figure}
\centering
\includegraphics[width=8cm]{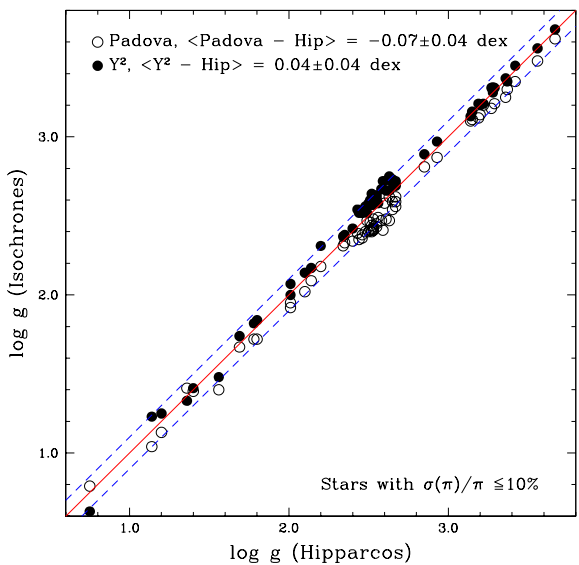}
\caption{Comparison between evolutionary gravities from $Y^2$ (filled circles)
and Padova (open circles) isochrones, and trigonometric gravities
for giant stars with good ($\sigma(\pi)/\pi \le$ 10\%) Hipparcos parallaxes.
The solid and dashed lines depict perfect agreement and variations of $\pm$0.1 dex,
respectively. }
\label{f:logg}
 \end{figure}

\begin{figure}
\centering
\includegraphics[width=8cm]{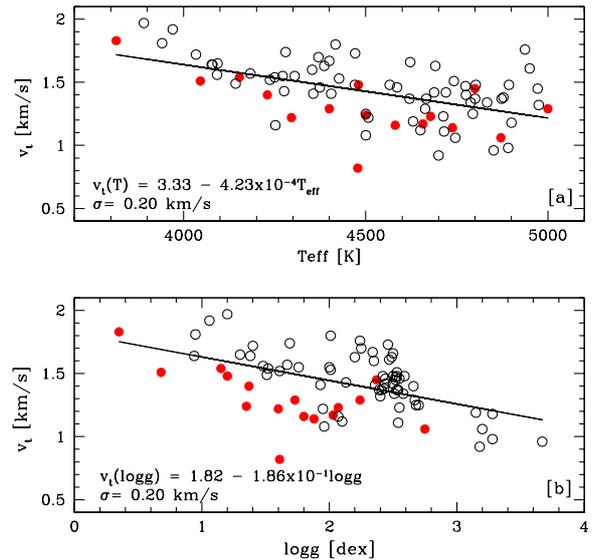}
\caption{Microturbulent velocity as a function of effective temperature ({\it
upper panel}) and log $g$ ({\it lower panel}).
Giant stars with [Fe/H] $<$ --0.70 and [Fe/H] $\geq$ --0.70 are, respectively,
represented by {\it filled} and {\it open} circles.
The solid line is a linear least squares fit to the data, whose results are
labeled in the figure.}
\label{f:vttemp}
\end{figure}


Photometric temperatures were obtained using optical and infrared colors and the 
infrared flux method $T_{\rm eff}$-scale of \citet{2005ApJ...626..465R}.
We note that the new improved IRFM calibration of
Casagrande et al. (2010) only applies to dwarf and subgiant stars
and thus cannot be applied to our sample of giants. However,
we do not expect any significant differences with respect to
\citet{2005ApJ...626..465R} for the relevant stellar parameters, 
except perhaps for a small ($\sim$1\%) zero-point offset in 
the \teff scale, which is irrelevant here since we are performing a
differential study.

Reddening for the bulge stars was estimated from extinction 
maps \citep{1996ApJ...460L..37S}, while for the comparison sample
both extinction maps \citep{mel06b} and Na\,{\sc i} D ISM 
absorption lines were used. The E(B-V) values based on the D lines were obtained as 
follows. In the optical thin case the relation between column density $N$ (units cm$^{-2}$)
and equivalent width $EW$ (units m\AA ) is:

\begin{equation}
N = 1.13 \times 10^{17} EW / (f \, \lambda^2 ), 
\end{equation}

\noindent  (Spitzer 1968).
The $f$ values are 0.64 and 0.32, respectively,
for the 5889.950 and 5895.924\,\AA\ lines 
(NIST database\footnote{http://physics.nist.gov/PhysRefData/ASD/lines\_form.html}). 
Note that the above 
relation between $N$(Na I) and equivalent width holds only for lines on the 
linear part of the curve of growth, i.e., for small values of E(B-V);
for reddening larger than a few 0.01\,mag the interstellar lines must be 
modeled in detail (e.g. Welty et al. 1994) to avoid underestimation of 
the column densities. In particular we use the profile fitting 
program FITS6P (Welty et al. 1994).

The $N$(Na I) density was transformed to $N$(H) using the relation found 
by Ferlet et al. (1985):

\begin{equation}
{\rm log} N({\rm H I + H_2})  = ({\rm log} N({\rm Na I}) + 9.09) / 1.04,
\end{equation}

\noindent where both $N$(Na I) and $N$(H) are in cm$^{-2}$.
Finally, E(B-V) was computed from the total hydrogen density (Bohlin et al.
1978):

\begin{equation}
E(B-V) = N({\rm H I + H_2}) / 5.8 \times 10^{21},
\end{equation}

\noindent where $N$(H) is in cm$^{-2}$ and E(B-V) in magnitudes. Although
this relation seems not well established for E(B-V)$<$0.1, Ramirez et al. (2006) have shown it to be very
accurate for a E(B-V)=0.01 star.

Albeit not used in the present work, we should mention for completeness that
in addition to reddening maps and Na D interstellar lines, 
E(B-V) can also be estimated from other interstellar 
features such as the diffuse interstellar band at 862 nm
(Munari et al. 2008),
as well as multicolor photometry \citep[e.g. Sect. 4.2 of][]
{mel06b,2006A&A...459..613R}
and polarization (e.g. Fosalba et al. 2002).

The stellar surface gravities were derived from improved Hipparcos 
parallaxes \citep{2007A&A...474..653V} for the sample of nearby giant stars and
assuming a distance of 8\,kpc for the bulge
giants. 
In addition, Yonsei-Yale \citep[$Y^2$;][]{2004ApJS..155..667D} and Padova 
isochrones \citep{2006A&A...458..609D} were employed to determine evolutionary 
gravities, as well as the input masses that were adopted
for the trigonometric gravities. In order to estimate the
$Y^2$ gravities, we generated a fine grid of isochrones, 
assuming [$\alpha$/Fe] = 0 and +0.3 for [Fe/H] $> 0$ and [Fe/H] $< -1$,
respectively, and linearly interpolated in between. All solutions
allowed by the error bars were searched for, adopting as final
result the median values.
The Padova gravities were obtained using the
Bayesian tool PARAM\footnote{http://stev.oapd.inaf.it/cgi-bin/param}.
As shown in Fig. \ref{f:logg}, both ($Y^2$, Padova) evolutionary gravities are 
in excellent agreement with the trigonometric gravities of our nearby giants
with reliable (uncertainties $\le$10\%) Hipparcos parallaxes.
The evolutionary $\log g$ values required 
small zero-point corrections of $-0.04$ ($Y^2$) and $+0.07$\,dex (Padova), 
to be on the same scale as the Hipparcos-based results for our sample 
giant stars (Fig. \ref{f:logg}). 
Bolometric corrections from \citet{1999A&AS..140..261A} were adopted.

We use iron lines to check our \teff and log $g$, but we do not assume a priori that our adopted effective temperatures, 
surface gravities, 1D model atmospheres, $gf$-values, selection of lines,
equivalent width measurements and LTE line formation, would result
in \emph{absolute} excitation (zero slope of Fe\,{\sc i} abundances vs. excitation potential)
and ionization ($A_{FeI} = A_{FeII}$) equilibria.
We use the nearby disk/halo giants to determine 
the slopes ($d(A_{FeI}) / d(\chi_{exc})$) and differences 
between Fe\,{\sc i} and Fe\,{\sc ii}, followed by most stars.
Our tests of the ionization and excitation balances of Fe\,{\sc i} and Fe\,{\sc ii} 
lines revealed that most of the sample giants (58 stars) satisfy our
\emph{relative} spectroscopic equilibrium of iron lines 
within the uncertainties, therefore 
the overall agreement is encouraging. Nevertheless, the photometric 
stellar parameters of 22 stars (8 thin disk, 5 thick disk, 1 halo, and 
8 bulge stars) required some adjustments to be on our \emph{relative} 
spectroscopic equilibrium scale. The corrections based on the trend 
followed by the bright disk/halo giants, for which the 
photometric stellar parameters (and stellar spectra) were more reliable than 
for the bulge sample, is:

\begin{equation}
d(A_{FeI}) / d(\chi_{exc}) = 0.008  \, {\rm dex \; eV^{-1} \;(Kurucz \; overshooting),}
\end{equation}

\begin{equation}
d(A_{FeI}) / d(\chi_{exc}) = 0.003  \, {\rm dex \; eV^{-1} \; (MARCS),}
\end{equation}

\noindent stars within 2-$\sigma$ ($\sigma$ = 0.011 dex eV$^{-1}$) 
were considered to fulfill our \emph{relative} excitation equilibrium.
Ionization balance was achieved if

\begin{equation}
A(\ion{Fe}{II}) - A(\ion{Fe}{I}) = 0.08 \, {\rm dex \;(Kurucz \; overshooting),}
\end{equation}

\begin{equation}
A(\ion{Fe}{II}) - A(\ion{Fe}{I}) = 0.00 \, {\rm dex \;(MARCS),}
\end{equation}

\noindent and stars within $\pm 0.07$\,dex were considered 
to fulfill our \emph{relative} ionization equilibrium.
After these corrections were performed the deviating thin disk, 
thick disk, halo and bulge stars, have stellar parameters in the same system,
i.e. in the Ram\'{\i}rez \& Mel\'endez (2005b) temperature scale and $\log g$ in the 
Hipparcos scale. 

The microturbulence was obtained by flattening any trend in 
the [FeI/H] versus reduced equivalent width diagram. 
The microturbulence follow tight relations with temperature and $\log g$
(Fig. \ref{f:vttemp}), with a scatter of only $0.20$\,km\,s$^{-1}$:

\begin{equation}
v_t(T_{\rm eff}) = 3.33 - 4.23 \times 10^{-4} T_{\rm eff} \hfill {\rm (MARCS)}
\end{equation}

\begin{equation}
v_t(T_{\rm eff}) = 3.40 - 4.41 \times 10^{-4} T_{\rm eff} \hfill {\rm (Kurucz \; overshooting)}
\end{equation}

\begin{equation}
v_t({\rm log} \, g) = 1.82 -0.186 \log g  \hfill {\rm (MARCS)}
\end{equation}

\begin{equation}
v_t({\rm log} \, g) = 1.84 -0.202 \log g  \hfill {\rm (Kurucz \; overshooting)}
\end{equation}

The lines used for analysis 
(presented as online material)
have been carefully selected to minimize the impact of blends. 
Completely avoiding blends is almost an impossible task in cool, relatively metal-rich giants
as in our sample, since their spectra are heavily blended
with many atomic and molecular lines (e.g. Coelho et al. 2005), in particular due to CN.
We have tried to avoid blending by performing spectral
synthesis of CN (using the line list of Mel\'endez \& Barbuy 1999)
and discarding the atomic lines whose equivalent widths are contaminated by
more than 10\% by CN. The cool giant Arcturus (Hinkle et al. 2000)
was also carefully inspected to discard lines that are severely
contaminated with other features. In some cases even lines which are
blended by more than 10\% have to be included, especially for elements
other than iron because only a few useful lines were available.
For heavily blended lines we have performed the measurements
by fitting only the unblended part of the profile, or 
deblending the feature using two or more components.
A preliminary version of our line list (Hekker \& Mel\'endez 2007) has 
been tested in $\sim$380 field (Hekker \& Mel\'endez 2007) and 39
open cluster (Santos et al. 2009) giants,
and the final list has been already used in field
bulge (Ryde et al. 2009b) and globular cluster (Mel\'endez \& Cohen 2009) giants.

The stellar chemical abundances were obtained from an equivalent width 
analysis using the 2002 version of  MOOG \citep{1973PhDT.......180S}. 
The same transition probabilities were applied to both the bulge and comparison samples. 
In the present work, we employed both Kurucz models with convective 
overshooting \citep{1997A&A...318..841C} and specially calculated 
MARCS (Gustafsson et al. 2008)
1D hydrostatic model atmospheres. For 
the MARCS models, both $\alpha$-enhanced ($[\alpha/{\rm Fe}]=+0.2$ and $+0.4$) 
and scaled-solar abundances models were constructed; for the Kurucz models, adjustments 
of [Fe/H] were applied to simulate the effects of $\alpha$-enhancement on the model 
atmospheres \citep{1993ApJ...414..580S}: 

\begin{equation}
\Delta  [{\rm Fe/H}] = \log (0.64 \times  10^{[\alpha/{\rm Fe}]} + 0.36)
\end{equation}

The effects of failing to account for the 
variations in $[\alpha/{\rm Fe}]$ are relatively small for a 
difference of $[\alpha/{\rm Fe}]$ = +0.1 dex (Table \ref{t:error}),
but could be important ($\sim$ 0.1 dex) for the typical enhancement 
of $[\alpha/{\rm Fe}]$ $\sim$ +0.3-0.4 seen in thick disk, halo and bulge stars. 

We estimate that our stellar parameters have typical 
uncertainties of 
$\Delta T_{\rm eff} \approx \pm 75$\,K, 
$\Delta \log g \approx \pm 0.3$\,dex and 
$\Delta v_{\rm t} \approx 0.2$\,km\,s$^{-1}$.
The impact of these uncertainties on the
abundance ratios, as well as the total abundance errors due to uncertainties
in \tsin, $\log g$, $v_{\rm t}$ and [$\alpha$/Fe] added in quadrature, are shown in
Table \ref{t:error}, but note that some uncertainties are likely to be
correlated to some degree (see, e.g., Fulbright et al. 2007). 
The uncertainties given in Table \ref{t:error} are probably conservative in some cases,
as shown by the relatively low scatter (as a function of metallicity)
of our abundance ratios; the uncertainties in the abundance
ratios [X/Fe] are probably $\leq$ 0.10 dex. 

{The uncertainty of $\pm$75K in \teff is the based on the upper and
lower envelopes (excluding outliers) of the differences between the slopes of iron abundance
vs. excitation potential ($d(A_{FeI}) / d(\chi_{exc})$) of the initial
photometric temperatures and the
adopted zero-points (relations 4 and 5). Note that these differences
are due not only to errors in the temperature calibrations, photometric
errors and the quality of the spectra, but also due to errors in E(B-V),
which although for the nearby giants are low, for the
bulge giants may be higher. Nevertheless,
since we correct all outliers from our adopted zero-points (which in some
cases may be due to incorrect reddenings), we are inmune to large errors in E(B-V).
Our error of 0.3 dex in log $g$ is based on the differences between FeII and FeI
from the initial trigonometric log $g$ and the adopted zero point
in FeII-FeI (relations 6 and 7). Note that since we are basing our
uncertainties on the upper and lower discrepancies of the initial
input stellar parameters and the adopted zero-points, our uncertainties
in \teff and log $g$ are conservative. For the bright disk/halo stars internal 
errors of 50K in \teff and 0.2 dex in log $g$ may be more adequate.
Ryde et al. (2009b) suggests that the uncertainties adopted in
the stellar parameters of our method (which was used to determine
the atmospheric parameters in their sample) are sound for their bulge giants. 
In particular, uncertainties in \teff higher than $\sim$75K are excluded 
based on the relatively low star-to-star scatter of their [O/Fe] ratios.

\begin{table}
\begin{flushleft}
\caption{Internal zero-point abundances adopted for our giant stars using MARCS and Kurucz models.}
\label{t:zero}      
\centering          
\begin{tabular}{lccccc}     
\noalign{\smallskip}
\hline\hline    
\noalign{\smallskip}
\noalign{\vskip 0.1cm} 
\multicolumn{1}{c}{\hbox{}} & \multicolumn{2}{c}{\hbox{\bf Sun}} & \multicolumn{2}{c}{\hbox{\bf Giants}} \\
Specie & Literature$^{\rm a}$ & This work$^{\rm b}$ & MARCS$^{\rm c}$ & Kurucz$^{\rm d}$  \\  
\noalign{\vskip 0.1cm}       
(1) & (2)  & (3) & (4) & (5) \\                    
\hline
\noalign{\smallskip}
\noalign{\vskip 0.1cm} 
Fe &  7.50,7.45,7.56,7.50 & 7.49$\pm0.04$ & 7.53 & 7.54 \\
O$_{\rm[O\,I]}$&  8.83,8.73.8.71,8.69 & 8.74$\pm0.04$ & 8.83 & 8.84 \\
Na &  6.33,6.27,6.27,6.24 & 6.24$\pm0.04$ & 6.24 & 6.24 \\
Mg &  7.58,7.54,7.58,7.60 & 7.56$\pm0.04$ & 7.65 & 7.66 \\
Al &  6.47,6.28,6.47,6.45 & 6.39$\pm0.04$ & 6.56 & 6.56 \\
Si &  7.55,7.62,7.54,7.51 & 7.54$\pm0.03$ & 7.60 & 7.63 \\
Ca &  6.36,6.33,6.36,6.34 & 6.34$\pm0.02$ & 6.32 & 6.30  \\
Ti &  5.02,4.90,4.92,4.95 & 4.94$\pm0.05$ & 4.83 & 4.81 \\
\hline                  
\end{tabular}
\begin{minipage}{.88\hsize}
 Notes.--- (a): Solar photospheric abundances from Grevesse \& Sauval (1998),
Reddy et al. (2003), Bensby et al. (2003, 2004) and Asplund et al. (2009);
(b): solar abundances based on our previous work 
(Mel\'endez et al. 2006a; Mel\'endez \& Ram\'{i}rez 2007; Mel\'endez et al. 2009) 
using different (e.g. McDonald, Keck, Magellan) solar spectra;
(c,d): Our internal zero-points for giants represent the thin-disk abundances
at [Fe/H] = 0.0. These zero-points are not absolute abundances and should 
only be used when both our same $gf$-values and analysis techniques are adopted \\
\end{minipage}			
\end{flushleft}
\end{table}

No predictions of the effects of 3D hydrodynamical models instead of classical
1D models used here are available as yet for the exact
stellar parameters of our targets \citep{2005ARA&A..43..481A}. 
\citet{2007A&A...469..687C} have performed such calculations for
slightly less evolved red giants ($T_{\rm eff} \approx 4700$ and $\log g \approx 2$)
and found that the the 3D abundance corrections 
for the species considered herein are expected to be modest: 
$|\Delta \log \epsilon| \la 0.1$\,dex at $[{\rm Fe/H}] \sim 0$.
At lower metallicity the 3D effects become more severe so that 
at [Fe/H]$=-1$ our 1D-based abundances could be in error by $\la 0.2$\,dex.
However, given the similarity in parameters between 
the bulge and disk giants, the {\em relative} abundance ratio differences -- which
we are primarily interested in here -- will be 
significantly smaller and thus inconsequential for our conclusions.

Of particular importance to our work are the adopted zero-points of our 
abundance scale. Most works use the Sun to define the zero-point of the thin
disk at [Fe/H] = 0.0, but due to the differences between dwarfs and giants,
this approach may introduce systematic errors. Instead, in the present work
we use seven thin disk giants with $-0.1 <$ [Fe/H] $<$+0.1 dex (HD 29503, HD 45415, 
HD 99648, HD 100920, HD 115478, HD 186378, HD 214376) to define our
zero points, which are shown in Table \ref{t:zero}
for both the Kurucz and MARCS models. 
In Table \ref{t:zero} we also show for comparison different abundance 
analysis of the Sun. As can be seen, our zero-points for Fe, Na and Ca 
are roughly in agreement with the solar abundances, but for O, Mg, and Si
the giants show a higher zero-point by $\sim$+0.1 dex, and for
Al differences as high as +0.15 are found. On the other hand,
Ti is lower by $\sim$0.1 dex. These zero-point offsets
of $-0.1$ to +0.15 dex show that it is not straightforward to
compare abundances obtained in giants with those found
in dwarfs.

These zero points we have found for giants are internal
for our particular set of $gf$-values and analysis techniques.
For comparison with chemical evolution models,
the absolute zero-points should be adopted from analysis of the Sun
(Asplund et al. 2009), which represents well the local thin disk 
at [Fe/H] = 0.0, except for small peculiarities of a few 0.01 dex
(Mel\'endez et al. 2009; Ram\'{i}rez et al. 2009).

The final stellar parameters are given in Table 4 and Table 5 for the MARCS
and the Kurucz models, respectively, while the abundance ratios are given
in Table 6 and Table 7. The equivalent width measurements are given in Tables 8-15,
which is available only in the electronic version of the article.
 
\addtocounter{table}{0} 
\addtocounter{table}{0} 
\addtocounter{table}{0} 
\addtocounter{table}{0} 

\begin{figure}
\centering
\includegraphics[width=8cm]{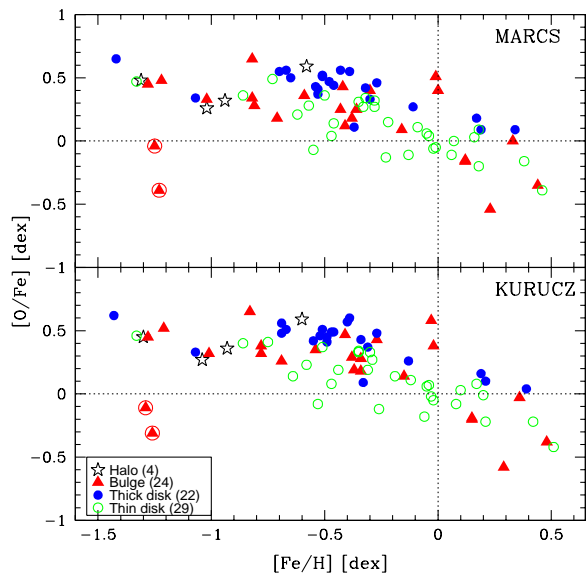}
\caption{[O/Fe] vs. [Fe/H] for the sample stars employing MARCS ({\it top})
and Kurucz ({\it bottom}) model atmospheres. Symbols are as explained in the
figure. Note, however, that hereafter
the bulge stars I-264 and IV-203 are omitted from all figures (refer to the text
for detail).}
\label{f:feo}
 \end{figure}
 
 \begin{figure}
\centering
\includegraphics[width=8cm]{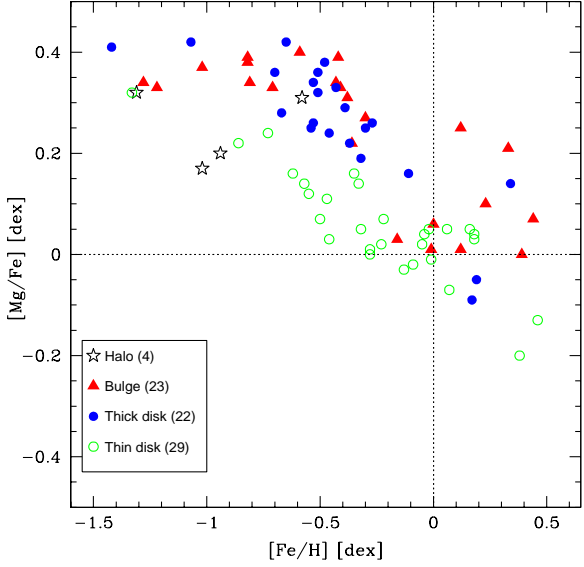}
\caption{[Mg/Fe] as a function of [Fe/H] for
MARCS model atmospheres. Symbols as explained in the figure.}
\label{f:femg}
 \end{figure}

\begin{figure}
\centering
\includegraphics[width=8cm]{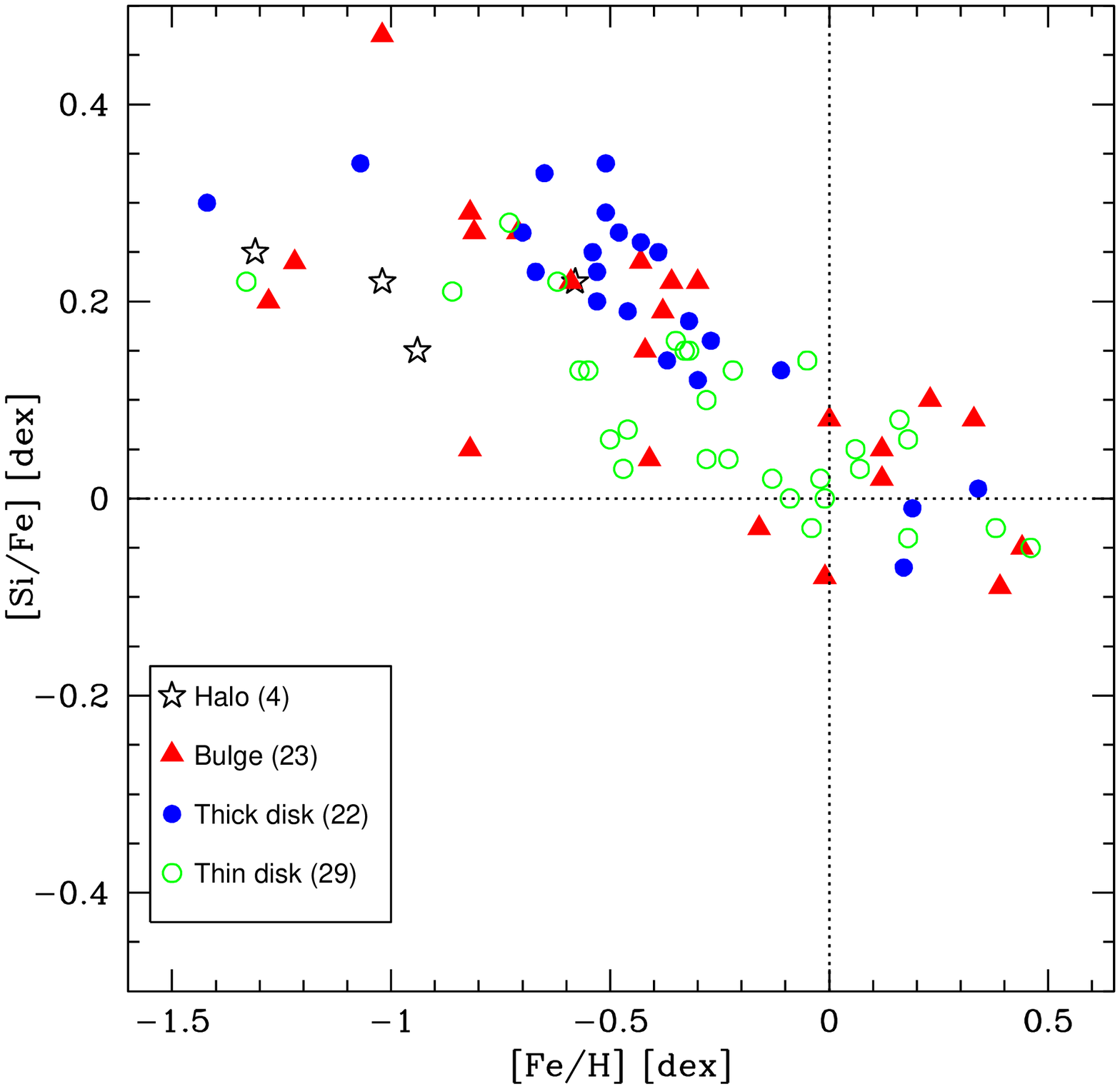}
\caption{[Si/Fe] vs. [Fe/H] for
MARCS model atmospheres.
Symbols as explained in the figure.}
\label{f:fesi}
 \end{figure} 
 
\begin{figure}
\centering
\includegraphics[width=8cm]{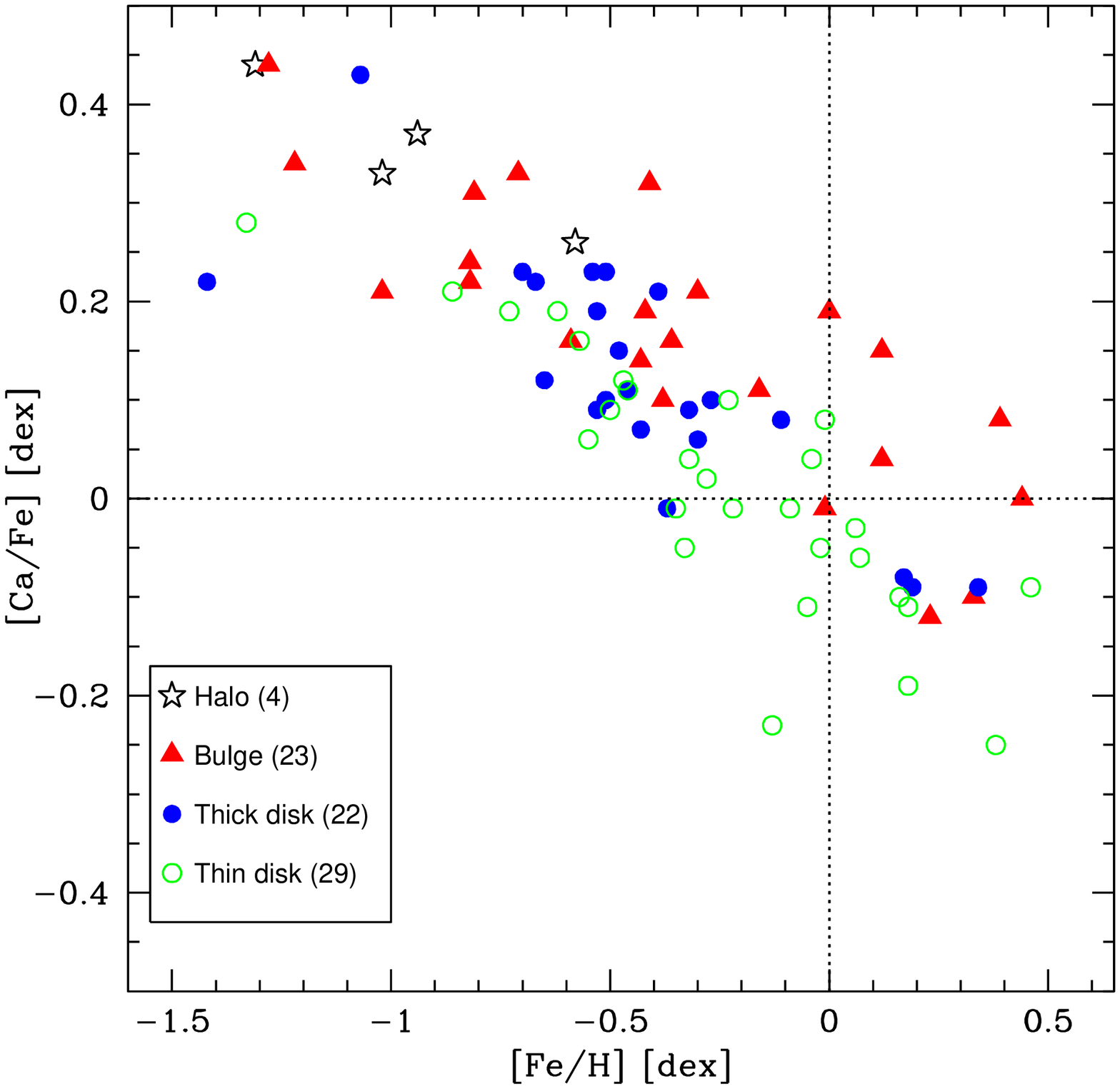}
\caption{[Ca/Fe] vs. [Fe/H]  for
MARCS model atmospheres.
Symbols as explained in the figure.}
\label{f:feca}
 \end{figure}
 
 \begin{figure}
\centering
\includegraphics[width=8cm]{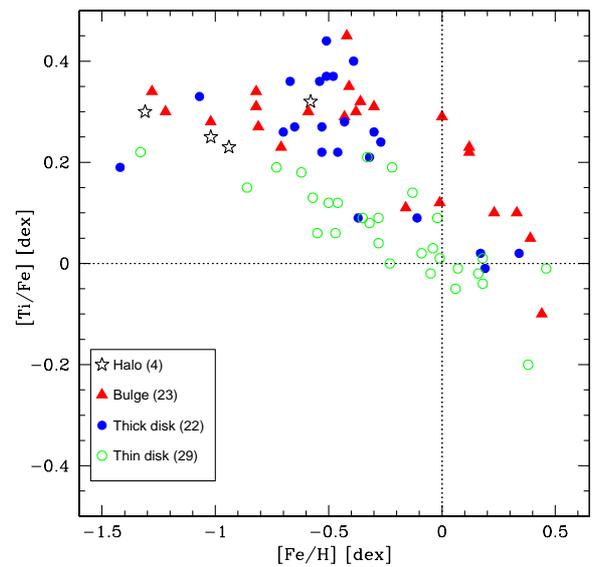}
\caption{Plot of [Ti/Fe] against [Fe/H] for the sample stars 
employing MARCS model atmospheres. Symbols are as explained in the figure.}
\label{f:feti}
 \end{figure}

\begin{figure}
\centering
\includegraphics[width=8cm]{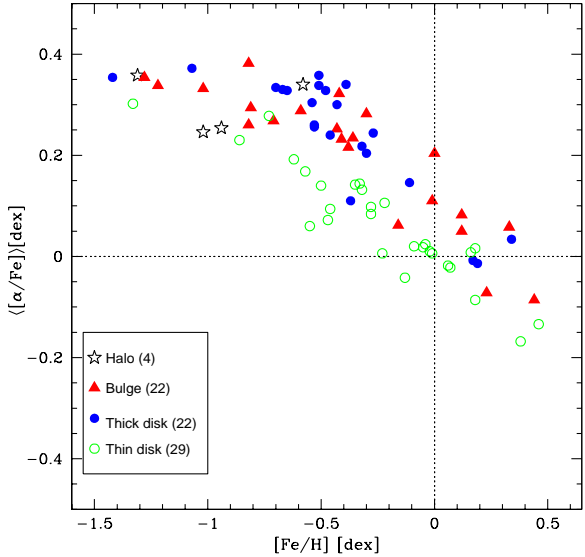}
\caption{Mean $\alpha$-elements abundance ratio ([(O,Mg,Si,Ca,Ti)/Fe]) as a
function of [Fe/H] for MARCS model atmospheres. 
Symbols as explained in the figure.}
\label{f:fealfaall}
\end{figure}

\begin{figure}
\centering
\includegraphics[width=8cm]{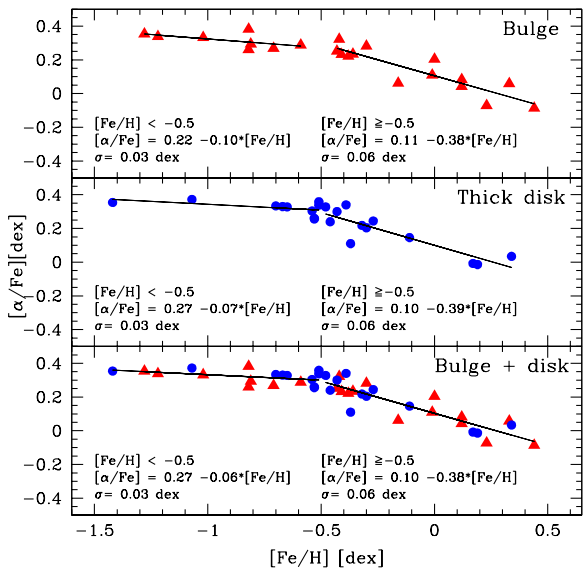}
\caption{Fit of [$\alpha$/Fe] vs. [Fe/H] for metal-poor
([Fe/H] $< -0.5$) and metal-rich ([Fe/H] $\ge -0.5$)
bulge ({\it top}), thick disk ({\it center}) 
and both bulge and thick disk ({\it bottom}) stars.
Both stellar populations can be fitted by similar relations,
with a scatter as low as $\sigma$ = 0.03 dex.
}
\label{f:fitalfa}
 \end{figure}

 \begin{figure}
\centering
\includegraphics[width=8cm]{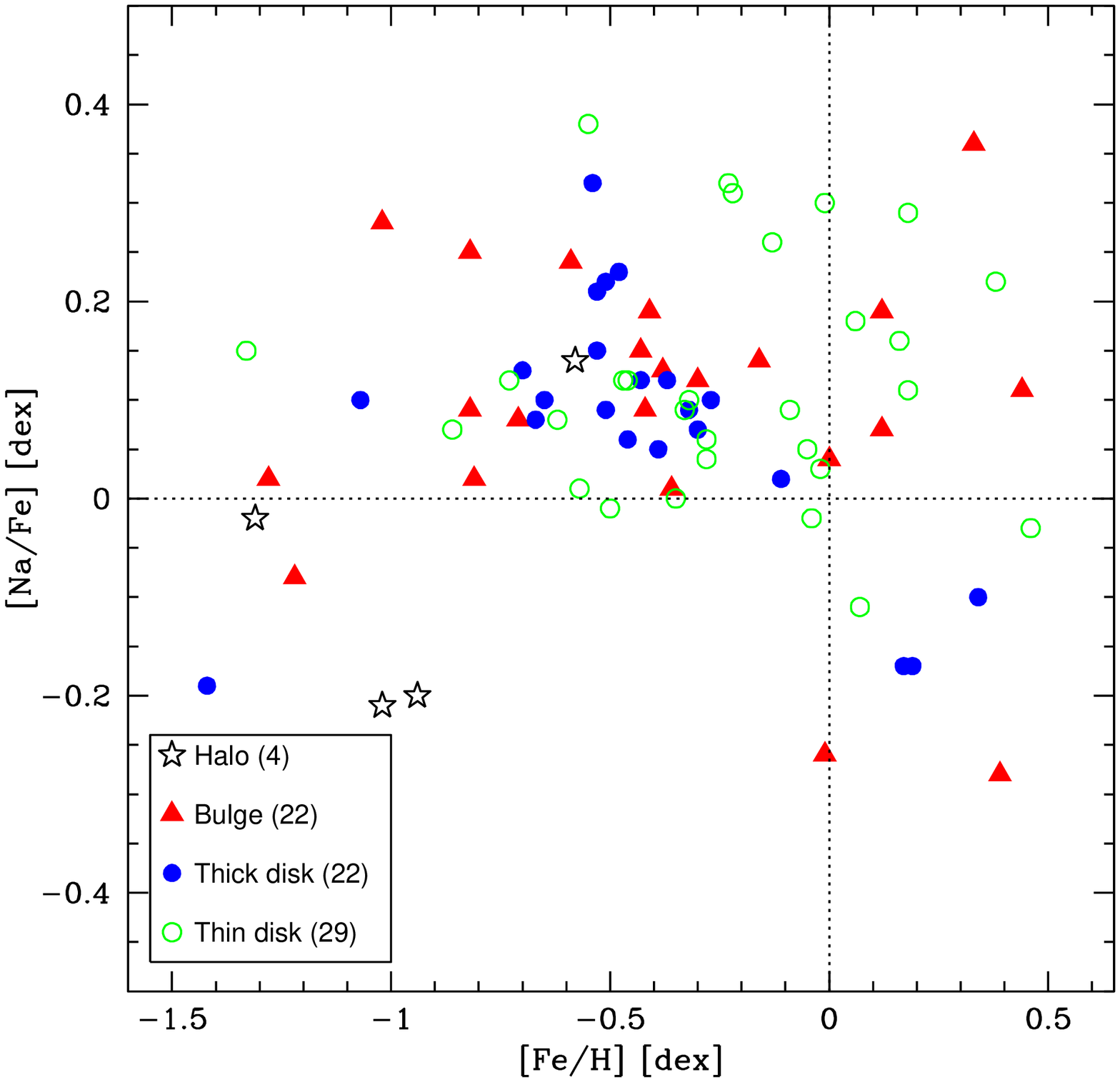}
\caption{[Na/Fe] vs. [Fe/H] for MARCS model atmospheres. Symbols as explained in the figure.}
\label{f:fena}
 \end{figure}

\begin{figure}
\centering
\includegraphics[width=8cm]{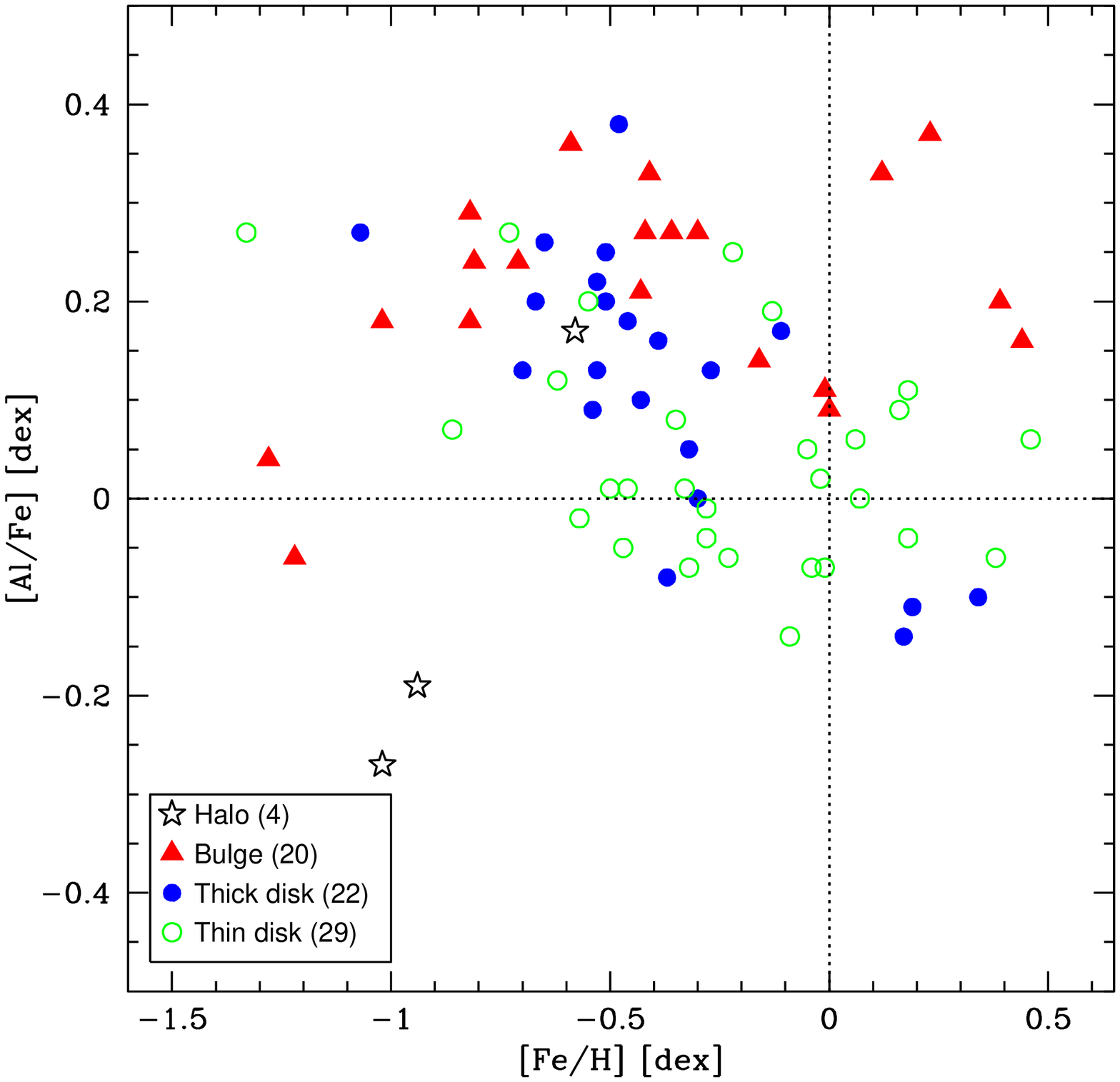}
\caption{[Al/Fe] vs. [Fe/H] for MARCS model atmospheres. 
Symbols as explained in the figure.}
\label{f:feal}
\end{figure}

\begin{figure}
\centering
\includegraphics[width=8cm]{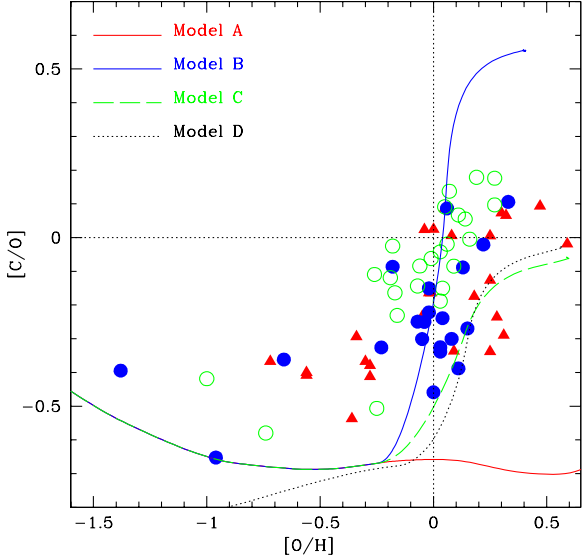}
\caption{[C/O] vs. [O/H] for the bulge ({\it triangles}), thick disk ({\it
filled circles}) and thin disk ({\it open circles}) with data taken from Mel\'endez et
al. (2008) and Ryde et al. (2009b). The original (undepleted) C abundances were
estimated from C+N (refer to the text). Overplotted, we show model predictions as
given in Cescutti et al. (2009; c.f. their Figure 5).}
\label{f:cohall}
 \end{figure} 
  
\begin{figure}
\centering
\includegraphics[width=8cm]{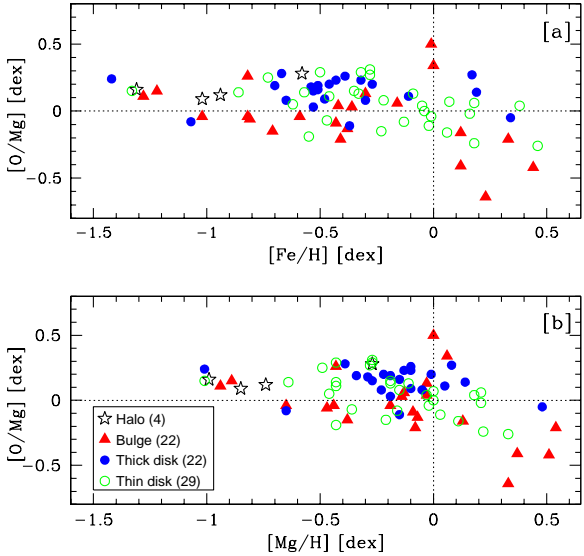}
\caption{[O/Mg] as a function of [Fe/H] ({\it top panel}) and [Mg/H] ({\it bottom panel}). 
Symbols as explained in the figure.}
\label{f:oxmg}
\end{figure}

\begin{figure}
\centering
\includegraphics[width=8cm]{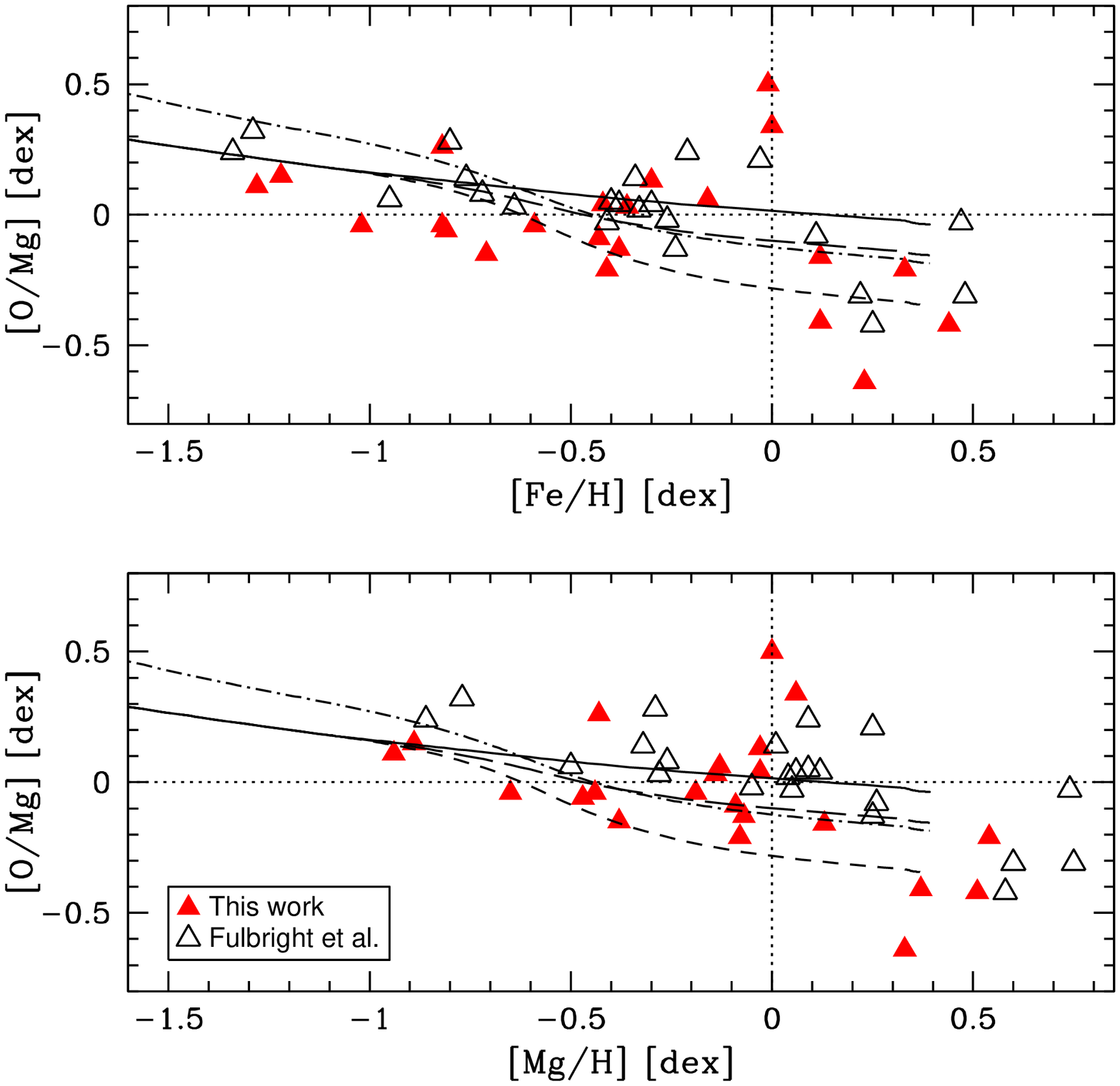}
\caption{[O/Mg] as a function of [Fe/H] for bulge giants 
according to Fulbright et al. (2007) (open triangles) and
our work (filled triangles). Both studies show a 
shallow trend up to about solar metallicity and a step
decrease in [O/Mg] for higher metallicities.
Recent predictions by Cescutti et al. (2009) are
shown as solid (WW95 model), short dashed (WW95+M92), 
long dashed (WW95+MM02) and dot-short dashed (MM02)
lines. All models have been shifted by $-0.2$ dex 
in [O/Mg] (see Cescutti et al. 2009 for a description of the models
and an explanation of the empirical offset).
}
\label{f:oxmgfulbrightmarcsshift}
\end{figure}
 
\section{Results}

In Fig. \ref{f:feo} we show the [O/Fe] ratios obtained in this work for
both MARCS and Kurucz overshooting model atmospheres.
As can be seen, there is a good
overall agreement between MARCS and Kurucz models. In particular, the
difference in iron abundance (MARCS -
Kurucz) is only $-$0.02 dex ($\sigma$ = 0.03 dex). Thus, in the following
figures, we present results based on the MARCS models only. 
Even though both set of models give similar chemical abundance ratios (see
Tables \ref{t:abundmarcs},\ref{t:abundkur}), for comparison with
chemical evolution models we suggest to adopt the MARCS results, since they were
computed with the correct [$\alpha$/Fe] ratio.
Our disk/halo comparison sample shows that the oxygen abundances
obtained here from the [O\,{\sc i}] 630 and 636\,nm lines 
and in Mel\'endez et al. (2008)
from infrared OH lines are in excellent agreement. 
The mean difference is 
only $-$0.04 dex (OH - [O\,{\sc i}]) 
with a scatter of 0.10 dex.
Since this is identical to the estimated error in [O/Fe] from OH found
in Mel\'endez et al. (2008), the uncertainties in our stellar parameters are
likely somewhat overestimated, as already discussed above.
Although the oxygen abundances obtained from [O\,{\sc i}] and OH lines
agree well, the results obtained from OH lines have less scatter, 
possibly because several OH lines were used instead
of relying on only one or two forbidden lines.
Therefore, for comparisons of oxygen abundances with detailed 
chemical evolution models of the thin disk, thick disk 
and bulge, we believe that the OH lines are preferable
(Mel\'endez et al. 2008; Ryde et al. 2009a,b). 
The [O\,{\sc i}]-based oxygen abundances confirms 
the similarity between the bulge and the local thick disk, 
which we previously demonstrated based on OH lines.
This is in contrast to some previous works on the topic
\citep{2006A&A...457L...1Z, 2007ApJ...661.1152F, 2007A&A...465..799L},
which argue that [O/Fe] in the bulge is higher than
in the thick disk based on a comparison to disk dwarf stars.

As pointed out by Fulbright et al. (2007), two O-deficient stars (I-264 and IV-203)
at $[{\rm Fe/H}] \approx -1.25$ have peculiar abundances similar to the 
O-Na anti-correlation seen in globular clusters 
(e.g. Gratton et al. 2004; Cohen \& Mel\'endez 2005; Yong et al. 2005;
Carretta et al. 2009).
Fulbright et al. (2007) found that these two giants have
high Na and Al abundances reminiscent of the
abundance anomalies seen in globular clusters, which we confirm.
Thus, the oxygen abundances of these two stars most likely
do not reflect the typical bulge composition. 

The results for the other $\alpha$-elements (Mg, Si, Ca, Ti) studied here
are shown in Figs. \ref{f:femg}-\ref{f:feti}.
As can be clearly seen, 
the chemical patterns of the bulge and thick disk are
indistinguishable also for those elements, reinforcing our previous findings 
based on oxygen abundances.
The average of our Mg abundances for the 7 stars with [Fe/H]$\ge$ 0 is [Mg/Fe] = 0.1$\pm$0.1,
in good agreement with the results from microlensed bulge dwarfs, which 
typically have [Mg/Fe]$\approx +0.1$
(Cohen et al. 2008, 2009; Johnson et al. 2008; Bensby et al. 2009a).
The latest preliminary results based on microlensed bulge dwarf stars 
(Bensby et al. 2009b) also indicate similarities between the 
bulge and thick disk for Ti and Mg at all 
probed metallicities ($-0.8 <$ [Fe/H] $< +0.5$).

Even more clear results are found when we combine
the results for all $\alpha$-elements, as shown in Fig. \ref{f:fealfaall}. 
It is clear that the chemical patterns of the bulge 
and thick disk are indistinguishable in their abundance patterns 
up to the metallicity range where the 
thick disk is unambiguously identified, i.e.,
up to $[{\rm Fe/H}]\approx -0.3$. Bensby et al. (2003, 2004)
reported the existence of a knee connecting thick-disk stars from [Fe/H]$\approx -0.3$ and
high [$\alpha$/Fe] to [Fe/H]$>$0.0 and low [$\alpha/$Fe], but 
Reddy et al. (2006) and Ram\'{i}rez et al. (2007) did not find any evidence of
such a knee. A re-examination of the latter
results including new observations of kinematically selected thick-disk
metal-rich objects is underway (Reddy et al., in preparation) and will
address this discrepancy. The problem is that at high [Fe/H] there is a
significant number of thick-disk candidates that follow the thin disk abundance
pattern.
That suggests that hot kinematics alone cannot be used to separate the thin from
the thick disk, especially at high [Fe/H]. At super-solar metallicities the problem may
be unsolvable because the abundance patterns of both disks merge.
Interestingly, the chemical similarities between the Galactic bulge and the local thick
disk giant stars we find in this work in fact extend to super-solar
metallicities. Yet, as explained above, it remains to be demonstrated that the few selected
thick disk stars are bona fide thick disk members rather than kinematically
heated thin disk stars.

From Table \ref{t:stars} we see that three giants, which 
are kinematically classified as thick disk (HD 77236, HD 107328) and thin
disk (HD 30608) members, present ambiguous kinematical population. The star
HD 77236 could be either a thick disk/halo star, while the stars HD 107328 and HD 30608
both have similar likelihood of belonging to the thin or thick disk
populations. These stars have [Fe/H] = ($-$0.67, $-$0.43, $-$0.28) and 
[$\alpha$/Fe] = (+0.33, +0.30, +0.08), respectively,
which means that both HD 77236 and HD 107328 could be indeed thick disk stars,
while the star HD 30608 has an abundance pattern consistent with 
a thin disk star at [Fe/H] $\sim -$0.3.

Linear fits of both the bulge and the thick disk [$\alpha$/Fe] vs. [Fe/H] 
relations up to [Fe/H] = $-0.3$, show that 
both populations follow identical patterns,
with a star-to-star scatter of only $\sigma$ = 0.03 dex.
We thus set the most stringent constraints to date 
on the chemical similarity of bulge and local thick disk stars.
The metallicity of the bulge extends to significantly higher [Fe/H] 
than that, which remains to be 
convincingly demonstrated for the thick disk, as previously discussed.

In order to quantify how similar are the bulge and the
thick disk at all metallicities, we have divided the stars
in a metal-poor and a metal-rich sample with the division set somewhat
arbitrarily at [Fe/H] = $-0.5$, and we have performed linear fits 
of [$\alpha$/Fe] vs. [Fe/H]. We find that the metal-poor part
of both stellar populations can be fitted by essentially identical relations
(Fig. \ref{f:fitalfa}): [$\alpha$/Fe] = 0.22 - 0.10 $\times$ [Fe/H]
for the bulge ($\sigma$ = 0.03 dex) and  [$\alpha$/Fe] = 0.27 - 0.07 $\times$ [Fe/H] 
for the thick disk ($\sigma$ = 0.03 dex).
These relations are identical to within $\pm$0.01 dex at [Fe/H] = $-1.5$
and to within $\pm$0.02 dex at [Fe/H] = $-0.5$, hence 
both datasets can be fitted by a single relation followed by both stellar populations:

\begin{equation}
{\rm [\alpha/Fe] = 0.268 -0.065 \times [Fe/H] \; \; ([Fe/H] < -0.5)},
\end{equation}

\noindent with a very low star-to-star scatter 
of only 0.03 dex (Fig. \ref{f:fitalfa}). 

A similar exercise for the most metal-rich bulge and thick disk
stars with [Fe/H] $\geq -0.5$, results in a single relation followed
by both stellar populations (Fig. \ref{f:fitalfa}):

\begin{equation}
{\rm [\alpha/Fe] = 0.104 -0.381 \times [Fe/H] \; \; ([Fe/H] \geq -0.5)},
\end{equation}

\noindent with a low star-to-star scatter 
of only 0.06 dex (Fig. \ref{f:fitalfa}). 
This scatter is higher than for the more metal-poor
stars ($\sigma$ = 0.03 dex), but fully explained by the higher uncertainties in the 
analysis of the more crowded spectra of the metal-rich stars.
Nevertheless, we do not discard that there may be some
contamination in the metal-rich samples.
We demonstrate thus that at all probed metallicities
($-1.5 <$ [Fe/H] $< +0.5$), both the bulge and thick disk
stellar populations have a striking chemical similarity.

The Al abundances of the most metal-rich bulge stars
([Fe/H] $\gtrsim$ 0) seem enhanced ([Al/Fe] $\sim$ +0.25), 
but Na seems solar (although with a large scatter) 
at these metallicities, as illustrated in Fig. \ref{f:fena}-\ref{f:feal}. Thus, the
enhancement in Al in metal-rich bulge giants is probably not
related to the Al-Na correlation seen in globular clusters, but most likely due to
the fact that the two Al\,{\sc i} lines employed are blended 
and are more difficult to deblend in the relatively
moderate S/N spectra of the metal-rich bulge giants. 
This is reinforced by
recent studies of metal-rich bulge dwarfs through
microlensing (Cohen et al. 2008; Johnson et al. 2008; Bensby et al. 2009a),
which find [Al/Fe] $\sim$ +0.10 dex. Thus, our
Al abundances for bulge stars with [Fe/H] $\gtrsim$ 0
are thus likely affected by systematic errors, which should be borne in mind in
comparisons with chemical evolution models.

\section{Discussion}

Our previous homogeneous abundance analysis of OH lines in high 
resolution infrared Gemini+Phoenix spectra 
was the first to show that the Galactic bulge and the local thick disk 
have indistinguishable [O/Fe] trends up to at least metallicities
[Fe/H] $\approx -0.3$ where the thick disk is unambiguously identified (Mel\'endez et al. 2008).
In the present work we have analyzed the forbidden oxygen lines 
and demonstrated that those indeed give consistent results with the OH lines.  
Importantly, we also extend the conclusions of
Mel\'endez et al. (2008) to other $\alpha$-elements (Mg, Si, Ca, Ti).
The $\alpha$-elements in the bulge and local thick disk stars 
are the same in the range $-1.5 <$ [Fe/H] $< -$0.3, 
showing a very low star-to-star scatter in [$\alpha$/Fe] 
of only 0.03 dex. 
Similarly, the [Na/Fe] and [Al/Fe] trends agree well for the bulge and
the local thick disk, although this is not too surprising given that there 
is no obvious offset between the thick and thin disk for those two elements.

Previous works (Fulbright et al. 2007; Lecureur et al. 2007) have found  
high [Mg/Fe] ratios in bulge stars, as well as high [X/Fe] ratios in other
elements with respect to the thick disk. It may seem surprising that using the
same equivalent widths as Fulbright et al. (2007) we find significantly lower [Mg/Fe] 
in bulge giants at all metallicities. However, as mentioned in the section 3, 
the zero-points we use in our analysis are based on seven thin disk solar 
metallicity giants, which, as shown in Table 3, are not the same as the
zero-points based on the Sun, which is $\sim$1400 K hotter and 
has a surface gravity $\sim$300 times higher than our giants.
The differences shown in Table 3 between the Sun and
our giants could be due to the different impact of 3D and non-LTE effects 
on giant and dwarfs, as well as problems with line blending in giants.
Furthermore, since we compare bulge giants to thick disk giants,
our conclusions on the similarity of the bulge and thick disk is
independent of the adopted zero-point, unlike the comparisons of
Zoccali et al. (2006), Fulbright et al. (2007) and Lecureur et al. (2007),
that compared bulge giants to disk dwarfs. Although both Fulbright et al. (2007)
and Lecureur et al. (2007) used the giants 
Arcturus ([Fe/H] $\sim -0.5$) and $\mu$ Leo ([Fe/H] $\sim$+0.3) 
as reference stars, their zero-points are ultimately based 
on the Sun\footnote{similar problems were already 
recognized 
by Gratton \& Sneden (1990) in their abundance analysis of  $\mu$ Leo, 
which 
was ultimately relative to the Sun since they used solar $gf$-values. 
They remark 
that their procedure might introduce inconsistencies due to the 
different atmospheric 
structures of $\mu$ Leo and the Sun. Furthermore, independently of the 
adopted 
$gf$-values, by definition the solar abundances are need to obtain 
[X/Fe].}, which as shown in Table 3, may be inadequate for the study of giants.
It is important to mention that Fulbright et al. (2007) also included 17 nearby 
disk giants (mostly from the thin disk) in order to check whether
they follow the same abundance pattern as disk dwarfs, but most of their
giants were observed at a lower resolving power, R $\sim$ 30,000,
implying higher errors in the abundances obtained from the
crowded spectra of metal-rich giant stars. Furthermore, their comparison sample
of disk dwarfs included several studies which may have different systematic
offsets between them. For example the study by Reddy et al. (2003) used
Str\"omgren photometry to estimate effective temperatures using the $(b-y)$ calibration
by Alonso et al. (1996). However, as recently shown by Mel\'endez et al. (2010,
in preparation) using $uvby$-$\beta$ photometry of solar twins, this calibration 
has a zero-point error of 130 K. In turn, this implies abundance variations
($\Delta$[X/H]) from -0.12 dex (O based on OI triplet,
which was the main abundance indicator used by Reddy et al. 2003) to +0.13 dex
(Ti). Yet, due to a compensating change in iron, for
most elements studied here (except for
[O/Fe] that is affected by -0.22 dex; based on the OI triplet), the
$\Delta$[X/Fe] ranges from -0.08 dex ([Si/Fe])
to +0.03 dex ([Ti/Fe]). On the other hand, Bensby et al. (2003, 2004) adopted
spectroscopic temperatures, therefore not only their Teff but also their [X/Fe]
abundance ratios are likely more accurate for comparison of these reddened
bulge regions.

Thus, the main reason why our conclusions regarding the abundance trends
in the bulge in comparison with the thick and thin disk differ from the 
findings of previous studies
(e.g. Zoccali et al. 2006; Fulbright et al. 2007; Lecureur et al. 2007)
is that we perform a strictly differential analysis of red giants 
with very similar parameters for all populations
rather than relying on either literature data or using dwarf stars for the disk samples.
We therefore bypass several potential systematic errors that
can scupper any analysis (e.g. 3D, non-LTE, stellar parameters, $gf$-values, 
see discussion in Asplund 2005). 
It is also worth noting that Bensby et al. (2009b) have shown that microlensed dwarfs
in the Galactic bulge present [$\alpha$/Fe] abundance ratios
similar to those of dwarfs in the Galactic thick disk, which
is in agreement with our results for giant stars.

The identical enhancement of the $\alpha$-elements in the bulge and
the local thick disk -- including the location of the knee that canonically 
is supposed to reflect the start of significant contribution of Fe production from
SNe\,Ia -- argues that the two stellar populations not only shared a
similar star formation rate but also initial mass function, in contrast
to the conclusions of some recent studies 
(e.g. Ballero et al. 2007; McWilliam et al. 2008; Cescutti et al. 2009).
We emphasize that this similarity does not automatically imply a casual connection
between the bulge and the local thick disk but it would be worthwhile
exploring this possibility further, in particular since such a relationship
has been proposed for other spiral galaxies
(e.g. van der Kruit \& Searle 1981). 
A worthwhile avenue to pursue would be the effects
of Galactic radial migration
(e.g. Sellwood \& Binney 2002; Haywood 2008; Roskar et al. 2008; Sch\"onrich \& Binney 2009a,b).
We note especially the hypothesis by Sch\"onrich \& Binney (2009b)
that the thick disk is a natural consequence of radial mixing, i.e. the thick disk
stars in the solar neighborhood originated in the inner part of the Galaxy.
It would therefore be particularly interesting to carry out a detailed chemical
analysis of a sample of in-situ inner disk stars to investigate any chemical
similarities with the bulge and the local thick disk giants studied herein.

McWilliam et al. (2008) have argued that their observed steadily
declining [O/Mg] vs. [Mg/H] trends for both the bulge and the solar
vicinity are the result of metallicity-dependent O yields due to mass-loss in
massive stars, possibly augmented by stellar rotation 
(e.g. Maeder 1992; Meynet \& Maeder 2002). 
Since these stellar winds remove C that would otherwise be converted to O,
the decreasing O production should be accompanied by enhanced C yields.
Cescutti et al. (2009) have extended the chemical modelling of
McWilliam et al. by investigating the resulting [C/O] trends. 
Although neither of their models match the observed [C/O] vs. [O/H] ratios
for the bulge or disk in detail, they argue that the observations support 
the metallicity-dependent yields of massive stars.

In Fig. \ref{f:cohall} we plot the [C/O] vs. [O/H] ratios of bulge and
disk stars based on Melendez et al. (2008) and Ryde et al. (2009b), which are both in similar
abundance scales. As in Cescutti et al. (2009), the primordial C abundances were estimated by
adding C+N and subtracting a ``primordial'' N abundance assuming [N/Fe]=0. As can be seen,
none of the Cescutti et al. models provide a good fit to the data. As mentioned by them, adopting
an IMF not as skewed to massive stars as that adopted by Ballero et al. (2007) may help to
alleviate the discrepancy, but on the other hand it may ruin their fit to the
bulge metallicity distribution.

Regarding the [O/Mg] ratios, our own observational data paint a somewhat different picture than the 
one presented by McWilliam et al. (2008) and Cescutti et al. (2009). 
Fig. \ref{f:oxmg} shows our [O/Mg] results against both [Fe/H] and [Mg/H]. 
All three populations -- bulge, thick and thin disk -- are similar up to
at least solar metallicity and follow an essentially flat trend with [O/Mg]$\approx 0.0-0.1$.
The similarities between the thick and thin disks 
extend even further provided the few stars with [Fe/H]$>0$ 
classified as belonging to the thick disk kinematically are truly bona-fide members.
There is some indication that for the bulge [O/Mg] becomes negative for super-solar
metallicities in line with the findings of Lecureur et al. (2007) and McWilliam et al. (2008).
They claim however an essentially continuous downward trend over the entire metallicity span
of their sample while we find a flat trend with a possible break around solar [Fe/H]. 
Nevertheless, a comparison between Fulbright et al. (2007) and our [O/Mg] ratios for bulge giants
(Fig. \ref{f:oxmgfulbrightmarcsshift}), shows that their [O/Mg] ratios actually do not have
a continuous downward trend, but instead 
their data show a shallow trend up to [Fe/H] $\sim +0.1$,
and then a step decrease for higher metallicities. 
Thus, both Fulbright et al. (2006, 2007) and our own analysis shows that there may be 
a break around solar metallicity in the [O/Mg] ratios.
However, none of the models presented in Cescutti et al. (2009)
shows the sharp break around solar metallicity 
indicated by the bulge giants (Fig. \ref{f:oxmgfulbrightmarcsshift}). 
The existence of this break would have to be confirmed by a significantly 
larger bulge sample than we have access to here; 
adding the most metal-rich stars from the Lecureur et al. study
would be worthwhile in this respect (as mentioned earlier, our bulge 
sample consists of the stars observed by Fulbright et al. 2006, 2007). 
If real, the downward trend could be a manifestation of 
metallicity-dependent O yields due to mass-loss in massive stars.

\begin{acknowledgements}
AAB acknowledges CAPES for financial support 4685-06-7 (PDE) and a FAPESP
fellowship no. 04/00287-9. We would like to thank the anonymous
referee for helpful constructive comments on the paper. Likewise, we are
grateful to A. McWilliam for sharing the Keck spectrum 
of one bulge star used for comparison purposes, and
G. Cescutti for sending the models shown in
Figs.\ref{f:cohall},\ref{f:oxmgfulbrightmarcsshift}.
This work has been supported by the
Australian Research Council (DP0588836), 
ANSTO (06/07-0-11), National Science Foundation 
(AST 06-46790), and the Portuguese FCT/MCTES (project PTDC/CTE-AST/65971/2006).
J.M. is supported by a Ciencia 2007 contract funded
by FCT/MCTES (Portugal) and POPH/FSE (EC).

Based partly on observations obtained at the
Las Campanas Magellan telescopes
(through Australian time with travel support provided by the Australian Access 
to Major Facilities Programme 06/07-O-11),
Keck Observatory (operated jointly by the California
Institute of Technology, the University of California and the National
Aeronautics and Space Administration) and McDonald Observatory.
\end{acknowledgements}


\begin{longtable}{lccccccc}
\caption{Final atmospheric parameters using the MARCS models}             
\label{t:atmosmarcs}   \\   
\hline\hline    
\noalign{\smallskip}
\noalign{\vskip 0.1cm} 
Star & T$_{\rm eff}$ [K] & log g [dex] & v$_{\rm t}$ [km/s] & [FeI/H] $\pm$ $\sigma$ & N &[FeII/H] $\pm$ $\sigma$ & N  \\        
(1) & (2)  & (3)  & (4)  & (5)  & (6) & (7)  & (8) \\                    
\hline 
\noalign{\vskip 0.1cm}
\endfirsthead

\multicolumn{8}{c}%
{{\tablename\ \thetable{} -- continued}} \\
\hline \hline 
\noalign{\smallskip}
\noalign{\vskip 0.1cm} 
Star & T$_{\rm eff}$ [K] & log g [dex] & v$_{\rm t}$ [km/s] & [FeI/H] $\pm$ $\sigma$ & N &[FeII/H] $\pm$ $\sigma$ & N  \\        
(1) & (2)  & (3)  & (4)  & (5)  & (6) & (7)  & (8) \\                    

\noalign{\vskip 0.1cm} 
\hline \\
\endhead
\endfoot
\hline
\endlastfoot
\noalign{\vskip 0.1cm}

\noalign{\vskip 0.1cm}    
\multicolumn{8}{c}{\hbox{\bf Halo}} \\
\noalign{\vskip 0.1cm}
   HD041667   & 4581   & 1.80  &    1.16  &   -1.02  $\pm$ 0.07  & 41  & -0.96 $\pm$ 0.08  & 6  \\
   HD078050   & 5000   & 2.24  &    1.29  &   -0.94  $\pm$ 0.07  & 36  & -0.87 $\pm$ 0.08  & 7  \\
   HD114095   & 4794   & 2.68  &    1.25  &   -0.58  $\pm$ 0.07  & 42  & -0.62 $\pm$ 0.06  & 7  \\
   HD210295   & 4738   & 1.88  &    1.14  &   -1.31  $\pm$ 0.07  & 35  & -1.35 $\pm$ 0.09  & 5  \\
\noalign{\vskip 0.1cm}    
\multicolumn{8}{c}{\hbox{\bf Bulge}} \\
\noalign{\vskip 0.1cm}
        I012  & 4237   & 1.61  &    1.52  &   -0.43 $\pm$ 0.08  & 26  & -0.44 $\pm$ 0.06  & 3  \\
        I025  & 4370   & 2.25  &    1.70  &    0.44 $\pm$ 0.10  & 19  &  0.39 $\pm$ 0.12  & 4  \\
\hline                  
\end{longtable}

\begin{longtable}{lccccccc}
\caption{Final atmospheric parameters using the Kurucz models}             
\label{t:atmoskur}   \\   
\hline\hline    
\noalign{\smallskip}
\noalign{\vskip 0.1cm} 
Star & T$_{\rm eff}$ [K] & log g [dex] & v$_{\rm t}$ [km/s] & [FeI/H] $\pm$ $\sigma$ & N &[FeII/H] $\pm$ $\sigma$ & N  \\        
(1) & (2)  & (3)  & (4)  & (5)  & (6) & (7)  & (8) \\                    
\hline 
\noalign{\vskip 0.1cm}
\endfirsthead

\multicolumn{8}{c}%
{{\tablename\ \thetable{} -- continued}} \\
\hline \hline 
\noalign{\smallskip}
\noalign{\vskip 0.1cm} 
Star & T$_{\rm eff}$ [K] & log g [dex] & v$_{\rm t}$ [km/s] & [FeI/H] $\pm$ $\sigma$ & N &[FeII/H] $\pm$ $\sigma$ & N  \\        
(1) & (2)  & (3)  & (4)  & (5)  & (6) & (7)  & (8) \\                    

\noalign{\vskip 0.1cm} 
\hline \\
\endhead
\endfoot
\hline
\endlastfoot
\noalign{\vskip 0.1cm}

\noalign{\vskip 0.1cm}    
\multicolumn{8}{c}{\hbox{\bf Halo}} \\
\noalign{\vskip 0.1cm}
    HD041667   &  4581 &  1.80	&   1.15  &  -1.04  $\pm$   0.07    &	 41  &  -0.95  $\pm$   0.09    &      6  \\
    HD078050   &  5000 &  2.24  &   1.22  &  -0.93  $\pm$   0.07    &	 36  &  -0.79  $\pm$   0.09    &      7  \\
    HD114095   &  4794 &  2.68  &   1.23  &   -0.60  $\pm$   0.07    &	 42  &  -0.62  $\pm$   0.06    &      7  \\
    HD210295   &  4738 &  1.88  &   1.11  &   -1.30  $\pm$   0.07    &	 35  &  -1.35  $\pm$   0.09    &      5  \\
\noalign{\vskip 0.1cm}    
\multicolumn{8}{c}{\hbox{\bf Bulge}} \\
\noalign{\vskip 0.1cm}
        I012   &  4237 &  1.61  &    1.52 &   -0.38 $\pm$    0.08   &	  26 &   -0.24 $\pm$	0.07   &       3 \\
        I025   &  4370 &  2.25  &    1.70 &    0.48 $\pm$    0.09   &	  19 &    0.53 $\pm$	0.11   &       4 \\
\hline                  
\end{longtable}

\setlength\tabcolsep{2.5pt}
\begin{longtable}{lccccccccccccccc}
\caption{Abundance ratios using the MARCS models}             
\label{t:abundmarcs}   \\   
\hline\hline    
\noalign{\smallskip}
\noalign{\vskip 0.05cm} 
Star & [O/Fe] & $\sigma_{\rm [O/Fe]}$ & [Na/Fe] & $\sigma_{\rm [Na/Fe]}$ & [Mg/Fe] & $\sigma_{\rm [Mg/Fe]}$ & [Al/Fe] & $\sigma_{\rm [Al/Fe]}$ & [Si/Fe] &
$\sigma_{\rm [Si/Fe]}$ & [Ca/Fe] & $\sigma_{\rm [Ca/Fe]}$ & [Ti/Fe] & $\sigma_{\rm [Ti/Fe]}$ \\      
(1) & (2)  & (3)  & (4)  & (5)  & (6) & (7)  & (8) & (9) & (10)  & (11)  & (12)  & (13)  & (14) & (15)  \\                      
\hline 
\noalign{\vskip 0.05cm}
\endfirsthead

\multicolumn{15}{c}%
{{\tablename\ \thetable{} -- continued}} \\
\hline \hline 
\noalign{\smallskip}
\noalign{\vskip 0.05cm} 
Star & [O/Fe] & $\sigma_{\rm [O/Fe]}$ & [Na/Fe] & $\sigma_{\rm [Na/Fe]}$ & [Mg/Fe] & $\sigma_{\rm [Mg/Fe]}$ & [Al/Fe] & $\sigma_{\rm [Al/Fe]}$ & [Si/Fe] &
$\sigma_{\rm [Si/Fe]}$ & [Ca/Fe] & $\sigma_{\rm [Ca/Fe]}$ & [Ti/Fe] & $\sigma_{\rm [Ti/Fe]}$ \\      
(1) & (2)  & (3)  & (4)  & (5)  & (6) & (7)  & (8) & (9) & (10)  & (11)  & (12)  & (13)  & (14) & (15)  \\             

\noalign{\vskip 0.05cm} 
\hline \\
\endhead
\endfoot
\hline
\endlastfoot
\noalign{\vskip 0.05cm}
\noalign{\vskip 0.05cm}    
\multicolumn{15}{c}{\hbox{\bf Halo}} \\
\noalign{\vskip 0.05cm}
    HD041667	&  0.26    &	 0.01  &    -0.21 &	 0.09   &     0.17    &     0.04  &    -0.27  &      0.12  &	 0.22  &      0.12   &    0.33   &	0.07  &      0.25   &	  0.08  \\
    HD078050	&  0.32    &	0.07   &    -0.20 &	 0.07	&     0.20    &    0.06   &    -0.19  &      ...   &	 0.15	&     0.07    &   0.37    &    0.05   &      0.23   &	 0.07	\\
    HD114095	&  0.59    &	 ...   &     0.14 &	 0.06	&     0.31    &    0.06   &     0.17  &     0.05   &	 0.22	&     0.10    &   0.26    &    0.08   &      0.32   &	 0.07	\\
    HD210295	&  0.48    &	0.01   &    -0.02 &	 0.09	&     0.32    &    0.07   &      ...  &      ...   &	 0.25	&     0.13    &   0.44    &    0.07   &       0.30   &	 0.08	\\
\noalign{\vskip 0.05cm}    
\multicolumn{15}{c}{\hbox{\bf Bulge}} \\
\noalign{\vskip 0.05cm}
        I012    &    0.25   &	  0.01  &      0.15 &	    0.19  &    0.34   &     0.11  &	  0.21  &      0.03  &     0.24  &	0.04   &     0.14   &	  0.06  &     0.29   &     0.09  \\
        I025    &   -0.35   &	  0.10  &      0.11 &	    0.22  &    0.07   &     0.10  &	  0.16  &      0.07  &    -0.05  &	0.04   &     0.00   &	  0.04  &    -0.10   &     0.09  \\
\hline                  
\end{longtable}

\setlength\tabcolsep{2.5pt}
\begin{longtable}{lccccccccccccccc}
\caption{Abundance ratios using the Kurucz models}             
\label{t:abundkur}   \\   
\hline\hline    
\noalign{\smallskip}
\noalign{\vskip 0.05cm} 
Star & [O/Fe] & $\sigma_{\rm [O/Fe]}$ & [Na/Fe] & $\sigma_{\rm [Na/Fe]}$ & [Mg/Fe] & $\sigma_{\rm [Mg/Fe]}$ & [Al/Fe] & $\sigma_{\rm [Al/Fe]}$ & [Si/Fe] &
$\sigma_{\rm [Si/Fe]}$ & [Ca/Fe] & $\sigma_{\rm [Ca/Fe]}$ & [Ti/Fe] & $\sigma_{\rm [Ti/Fe]}$ \\      
(1) & (2)  & (3)  & (4)  & (5)  & (6) & (7)  & (8) & (9) & (10)  & (11)  & (12)  & (13)  & (14) & (15)  \\                      
\hline 
\noalign{\vskip 0.05cm}
\endfirsthead

\multicolumn{15}{c}%
{{\tablename\ \thetable{} -- continued}} \\
\hline \hline 
\noalign{\smallskip}
\noalign{\vskip 0.05cm} 
Star & [O/Fe] & $\sigma_{\rm [O/Fe]}$ & [Na/Fe] & $\sigma_{\rm [Na/Fe]}$ & [Mg/Fe] & $\sigma_{\rm [Mg/Fe]}$ & [Al/Fe] & $\sigma_{\rm [Al/Fe]}$ & [Si/Fe] &
$\sigma_{\rm [Si/Fe]}$ & [Ca/Fe] & $\sigma_{\rm [Ca/Fe]}$ & [Ti/Fe] & $\sigma_{\rm [Ti/Fe]}$ \\      
(1) & (2)  & (3)  & (4)  & (5)  & (6) & (7)  & (8) & (9) & (10)  & (11)  & (12)  & (13)  & (14) & (15)  \\             

\noalign{\vskip 0.05cm} 
\hline \\
\endhead
\endfoot
\hline
\endlastfoot
\noalign{\vskip 0.05cm}
\noalign{\vskip 0.05cm}    
\multicolumn{15}{c}{\hbox{\bf Halo}} \\
\noalign{\vskip 0.05cm}
    HD041667	&    0.27   &	  0.01  &    -0.18   &     0.09   &	0.19	&    0.05    &   -0.24    &    0.12   &     0.22   &	 0.12	&     0.37   &     0.07    &	0.24   &    0.08   \\
    HD078050	&    0.36   &	  0.07  &    -0.19   &     0.07   &	0.22	&    0.04    &   -0.19    &	...   &     0.14   &	 0.07	&     0.39   &     0.05    &	0.25   &    0.07   \\
    HD114095	&    0.59   &	  ...	&     0.15   &     0.06   &	0.31	&    0.06    &    0.19    &    0.05   &     0.22   &	 0.10	&     0.29   &     0.08    &	0.34   &    0.07   \\
    HD210295	&    0.45   &	  0.01  &     0.00   &     0.09   &	0.33	&    0.07    &     ...    &	...   &     0.24   &	 0.14	&     0.48   &     0.07    &	0.33   &    0.08   \\
 \noalign{\vskip 0.05cm}
\noalign{\vskip 0.05cm}    
\multicolumn{15}{c}{\hbox{\bf Bulge}} \\
\noalign{\vskip 0.05cm}
        I012    &    0.29   &	  0.01  &      0.09  &      0.18  &	0.33   &     0.12   &	  0.14   &     0.03  &     0.27  &	0.04  &      0.08  &	  0.07   &    0.21  &	  0.09   \\
        I025    &   -0.38   &	  0.09  &      0.04  &      0.24  &	0.07   &     0.11   &	  0.14   &     0.07  &    -0.04  &	0.04  &     -0.02  &	  0.06   &   -0.13  &	  0.09   \\
\hline                  
\end{longtable}

\Online

\scriptsize

\onllongtab{8}{
\begin{longtable}{lcccccccccccccc}
\caption{Line list and measured equivalent widths given in m$\rm \AA$}             
\label{t:ew1}   \\   
\hline\hline    
\noalign{\smallskip}
\noalign{\vskip 0.05cm} 
$\lambda$ [$\rm \AA$]& $\chi$ [eV] & log$gf$ & HD041667 & HD078050 & HD114095 & HD210295  & I012  & I025  & I039  & I141  & I151  & I152  & I156  & I158   \\ 
 (1)      & (2)    & (3)     & (4)      & (5)      & (6)      & (7)       & (8)   & (9)   & (10)  & (11)  & (12)  & (13)  & (14)  & (15)  \\			  
\hline 
\noalign{\vskip 0.05cm}
\endfirsthead

\multicolumn{15}{c}%
{{\tablename\ \thetable{} -- continued}} \\
\hline \hline 
\noalign{\smallskip}
\noalign{\vskip 0.05cm} 
$\lambda$ [$\rm \AA$]& $\chi$ [eV] & log$gf$ & HD041667 & HD078050 & HD114095 & HD210295  & I012  & I025  & I039  & I141  & I151  & I152  & I156  & I158  \\ 
 (1)      & (2)    & (3)     & (4)      & (5)      & (6)      & (7)       & (8)   & (9)   & (10)  & (11)  & (12)  & (13)  & (14)  & (15) \\			 
\noalign{\vskip 0.05cm} 
\hline \\
\endhead
\endfoot
\hline
\endlastfoot
\noalign{\vskip 0.05cm}
{\bf FeI} &        &          &          &       &        &          &           &       &         &         &         &         &         &          \\  
5560.21   &   4.43 &   -1.120 &     41.3 &  29.1 &  48.6  &   23.2   &	  73.0   & 109.0 &    96.0 &	69.0 &	  47.0 &    79.0 &    56.0 &	79.0  \\
5633.95   &   4.99 &   -0.180 &    ...   &  41.6 &  59.3  &   29.3   &	  80.0   & 136.0 &     ... &	81.0 &	  56.0 &     ... &    62.0 &	94.0  \\
5638.26   &   4.22 &   -0.800 &     64.3 &  61.5 &  76.9  &   47.8   &	 105.0   &   ... &     ... &   100.0 &	   ... &   110.0 &    84.0 &   103.0  \\
5651.47   &   4.47 &   -1.850 &      9.5 &   ... &  18.6  &    5.3   &	  36.0   &  76.0 &    65.0 &	40.0 &	  21.0 &    48.0 &    23.0 &	44.0  \\
5662.52   &   4.18 &   -0.573 &     81.6 &  73.4 &  98.3  &   62.7   &	   ...   &   ... &     ... &   119.0 &	  91.0 &   128.0 &    96.0 &   121.0  \\
5679.02   &   4.65 &   -0.810 &     40.3 &  36.0 &  54.0  &   26.7   &	   ...   & 115.0 &    99.0 &	81.0 &	  51.0 &     ... &    50.0 &	76.0  \\
5731.76   &   4.26 &   -1.110 &     45.7 &  39.4 &  57.8  &   30.2   &	  86.0   & 129.0 &     ... &	81.0 &	  65.0 &    81.0 &    67.0 &	86.0  \\
\hline                  
\end{longtable}
} 

\onllongtab{9}{
\setlength\tabcolsep{1.3pt}
\begin{longtable}{lccccccccccccccccccc}
\caption{Line list and measured equivalent widths given in m$\rm \AA$}             
\label{t:ew2}   \\   
\hline\hline    
\noalign{\smallskip}
\noalign{\vskip 0.05cm} 
$\lambda$ [$\rm \AA$]& $\chi$ [eV] & log$gf$ & I194     & I202     & I264     & I322      & II033  & II119  & II154  & II172  & III152  & III220  & IV003  & IV047 & IV072  & IV167  & IV325  & IV329  & IV203  \\ 
 (1)      & (2)    & (3)     & (4)      & (5)      & (6)      & (7)       & (8)    & (9)    & (10)   & (11)   & (12)    & (13)    & (14)   & (15)  & (16)   & (17)   & (18)   & (19)   & (20)  \\			  
\hline 
\noalign{\vskip 0.05cm}
\endfirsthead

\multicolumn{20}{c}%
{{\tablename\ \thetable{} -- continued}} \\
\hline \hline 
\noalign{\smallskip}
\noalign{\vskip 0.05cm} 
$\lambda$ [$\rm \AA$]& $\chi$ [eV] & log$gf$ & I194     & I202     & I264     & I322      & II033  & II119  & II154  & II172  & III152  & III220  & IV003  & IV047 & IV072  & IV167  & IV325  & IV329  & IV203  \\ 
 (1)      & (2)    & (3)     & (4)      & (5)      & (6)      & (7)       & (8)    & (9)    & (10)   & (11)   & (12)    & (13)    & (14)   & (15)  & (16)   & (17)   & (18)   & (19)   & (20)  \\			  
\noalign{\vskip 0.05cm} 
\hline \\
\endhead
\endfoot
\hline
\endlastfoot
\noalign{\vskip 0.05cm}
{\bf FeI} &	   &	      & 	&	  &	    &	      & 	&	 &	    &	     &         &	 &	   &	     &         &	&	   &	     &  	\\  
5560.21   &   4.43 &   -1.120 &     ... &    76.0 &    48.0 &	 89.0 &    60.0 &  33.0  &   51.0   &  58.0  &   66.0  &   68.0  &   32.0  &   63.0  &   86.0  &   90.0 &     ...  &   52.0  &   43.1	\\
5633.95   &   4.99 &   -0.180 &    94.0 &    97.0 &    52.0 &	 95.0 &    67.0 &  38.0  &   56.0   &  70.0  &   78.0  &   76.0  &   40.0  &   88.0  &  103.0  &  121.0 &   110.0  &	...  &   52.4	\\
5638.26   &   4.22 &   -0.800 &     ... &   112.0 &    78.0 &	118.0 &    86.0 &  55.0  &   74.0   &  86.0  &   92.0  &   97.0  &   59.0  &   97.0  &    ...  &    ... &     ...  &	...  &   84.1	\\
5651.47   &   4.47 &   -1.850 &    39.0 &    52.0 &    17.0 &	 53.0 &    20.0 &   ...  &    ...   &  33.0  &   31.0  &   37.0  &    ...  &   32.0  &   51.0  &   62.0 &    52.0  &	...  &   13.4	\\
5662.52   &   4.18 &   -0.573 &   122.0 &     ... &    95.0 &	  ... &   102.0 &  74.0  &   89.0   & 100.0  &  117.0  &  108.0  &    ...  &  113.0  &    ...  &    ... &     ...  &	...  &  107.0	\\
5679.02   &   4.65 &   -0.810 &    77.0 &    79.0 &    45.0 &	 93.0 &    62.0 &   ...  &   56.0   &	...  &   76.0  &   73.0  &   40.0  &   79.0  &   86.0  &    ... &    97.0  &	...  &   52.1	\\
5731.76   &   4.26 &   -1.110 &     ... &    88.0 &    61.0 &	102.0 &    72.0 &  37.0  &   57.0   &  72.0  &   78.0  &   74.0  &   40.0  &   82.0  &    ...  &  109.0 &   117.0  &   64.0  &   63.0	\\
\hline                  
\end{longtable}
} 

\onllongtab{10}{
\setlength\tabcolsep{2.5pt}
\begin{longtable}{lccccccccccc}

\caption{Line list and measured equivalent widths given in m$\rm \AA$}             
\label{t:ew3}   \\   
\hline\hline    
\noalign{\smallskip}
\noalign{\vskip 0.05cm} 
$\lambda$ [$\rm \AA$]& $\chi$ [eV] & log$gf$ & HD077236 & HD023940 & HD032440 & HD037763  & HD040409 & HD077729 & HD080811 & HD083212 & HD099978    \\
 (1)      & (2)    & (3)     & (4)      & (5)      & (6)      & (7)       & (8)      & (9)      & (10)     & (11)     &   (12)      \\		      
\hline 
\noalign{\vskip 0.05cm}
\endfirsthead

\multicolumn{12}{c}%
{{\tablename\ \thetable{} -- continued}} \\
\hline \hline 
\noalign{\smallskip}
\noalign{\vskip 0.05cm} 
$\lambda$ [$\rm \AA$]& $\chi$ [eV] & log$gf$ & HD077236 & HD023940 & HD032440 & HD037763  & HD040409 & HD077729 & HD080811 & HD083212 & HD099978 \\
 (1)      & (2)    & (3)     & (4)      & (5)      & (6)      & (7)       & (8)      & (9)      & (10)     & (11)     &   (12)   \\			   
\noalign{\vskip 0.05cm} 
\hline \\
\endhead
\endfoot
\hline
\endlastfoot
\noalign{\vskip 0.05cm}

{\bf FeI} &	   &	      & 	&	  &	    &	     &       &         &	 &	    &		\\  
5560.21   &   4.43 &   -1.120 &   56.4  &   63.5  &   81.2  &	81.9 &  75.5 &    66.9 &    49.7 &    25.5  &	35.4	\\
5633.95   &   4.99 &   -0.180 &   67.0  &   81.2  &  102.5  &	92.3 &  95.6 &    81.1 &    63.1 &    33.5  &	43.7	\\
5638.26   &   4.22 &   -0.800 &   87.2  &   94.7  &  127.9  &  122.4 & 110.2 &   103.4 &    73.5 &    52.7  &	64.2	\\
5651.47   &   4.47 &   -1.850 &   26.7  &   29.4  &   50.6  &	52.2 &  44.7 &    37.0 &    17.6 &     ...  &	10.3	\\
5662.52   &   4.18 &   -0.573 &  105.1  &  107.7  &  136.0  &  136.0 & 126.1 &   120.1 &   100.6 &    77.1  &	79.4	\\
5679.02   &   4.65 &   -0.810 &   63.2  &   77.5  &   86.1  &	87.9 &  84.0 &    67.4 &    54.9 &     ...  &	41.0	\\
5731.76   &   4.26 &   -1.110 &   71.1  &   75.3  &  101.3  &  101.7 &  89.4 &    84.5 &    57.9 &     ...  &	43.9	\\
\hline                  
\end{longtable}
} 

\onllongtab{11}{
\setlength\tabcolsep{2.5pt}
\begin{longtable}{lccccccccccc}

\caption{Line list and measured equivalent widths given in m$\rm \AA$}             
\label{t:ew4}   \\   
\hline\hline    
\noalign{\smallskip}
\noalign{\vskip 0.05cm} 
$\lambda$ [$\rm \AA$]& $\chi$ [eV] & log$gf$ & HD107328 & HD107773 & HD119971 & HD124897  & HD127243 & HD130952 & HD136014 & HD145148 & HD148451    \\
 (1)      & (2)    & (3)     & (4)      & (5)      & (6)      & (7)       & (8)      & (9)      & (10)     & (11)     &   (12)      \\		      
\hline 
\noalign{\vskip 0.05cm}
\endfirsthead

\multicolumn{12}{c}%
{{\tablename\ \thetable{} -- continued}} \\
\hline \hline 
\noalign{\smallskip}
\noalign{\vskip 0.05cm} 
$\lambda$ [$\rm \AA$]& $\chi$ [eV] & log$gf$ & HD107328 & HD107773 & HD119971 & HD124897  & HD127243 & HD130952 & HD136014 & HD145148 & HD148451    \\
 (1)      & (2)    & (3)     & (4)      & (5)      & (6)      & (7)       & (8)      & (9)      & (10)     & (11)     &   (12)   \\			   
\noalign{\vskip 0.05cm} 
\hline \\
\endhead
\endfoot
\hline
\endlastfoot
\noalign{\vskip 0.05cm}

{\bf FeI} &	   &	      & 	&	  &	    &	     &       &        &       &        &	 \\  
5560.21   &   4.43 &   -1.120 &    71.2 &   52.3  &  66.1   & 72.5   & 47.2  &  65.1  &  60.0 &  79.1  &  50.0    \\
5633.95   &   4.99 &   -0.180 &    89.9 &   63.4  &  73.0   & 84.0   & 59.3  &  81.3  &  74.2 &  94.8  &  59.0    \\
5638.26   &   4.22 &   -0.800 &   111.5 &   76.8  &  97.9   &105.0   & 78.0  &  92.2  &  89.0 & 107.2  &  78.1    \\
5651.47   &   4.47 &   -1.850 &    36.6 &   22.0  &  33.2   & 34.0   & 15.4  &  30.8  &  24.3 &  37.8  &  18.9    \\
5662.52   &   4.18 &   -0.573 &   126.7 &   94.1  & 121.3   &123.0   & 91.1  & 118.9  & 106.3 & 127.4  &  94.7    \\
5679.02   &   4.65 &   -0.810 &    78.7 &   59.1  &  68.4   & 76.0   & 51.8  &  68.5  &  66.9 &  80.0  &  55.1    \\
5731.76   &   4.26 &   -1.110 &    85.5 &   62.6  &  82.7   & 91.0   & 54.2  &  81.8  &  72.1 &  83.7  &  60.0    \\
\hline                  
\end{longtable}
} 

\onllongtab{12}{
\setlength\tabcolsep{2.5pt}
\begin{longtable}{lccccccccccc}

\caption{Line list and measured equivalent widths given in m$\rm \AA$}             
\label{t:ew5}   \\   
\hline\hline    
\noalign{\smallskip}
\noalign{\vskip 0.05cm} 
$\lambda$ [$\rm \AA$]& $\chi$ [eV] & log$gf$ & HD180928 & HD203344 & HD219615 & HD221345  & HD000787 & HD003546 & HD005268 & HD029139 & HD029503   \\
 (1)      & (2)    & (3)     & (4)      & (5)      & (6)      & (7)       & (8)      & (9)      & (10)     & (11)     &   (12)      \\		      
\hline 
\noalign{\vskip 0.05cm}
\endfirsthead

\multicolumn{12}{c}%
{{\tablename\ \thetable{} -- continued}} \\
\hline \hline 
\noalign{\smallskip}
\noalign{\vskip 0.05cm} 
$\lambda$ [$\rm \AA$]& $\chi$ [eV] & log$gf$ & HD180928 & HD203344 & HD219615 & HD221345  & HD000787 & HD003546 & HD005268 & HD029139 & HD029503 \\
 (1)      & (2)    & (3)     & (4)      & (5)      & (6)      & (7)       & (8)      & (9)      & (10)     & (11)     &   (12)   \\			   
\noalign{\vskip 0.05cm} 
\hline \\
\endhead
\endfoot
\hline
\endlastfoot
\noalign{\vskip 0.05cm}

{\bf FeI} &	   &	      & 	   & 	      &          &          &            &          &            &          &         \\  
5560.21   &   4.43 &   -1.120 &    69.7    &   74.5   &   58.5   &    71.6  &     90.0   &    50.4  &     57.3   &    91.0  &     80.4 \\
5633.95   &   4.99 &   -0.180 &    79.8    &   92.0   &   66.8   &    85.4  &    109.5   &    64.0  &     68.6   &     ...  &     92.6 \\
5638.26   &   4.22 &   -0.800 &   102.2    &  106.0   &   83.5   &   100.4  &    137.0   &    80.1  &     84.1   &     ...  &    111.5 \\
5651.47   &   4.47 &   -1.850 &    38.4    &   37.4   &   19.7   &    32.9  &     53.4   &    16.8  &     21.4   &     ...  &     49.1 \\
5662.52   &   4.18 &   -0.573 &   117.5    &  114.7   &   98.3   &   110.0  &    140.7   &    91.4  &    100.7   &     ...  &    135.8 \\
5679.02   &   4.65 &   -0.810 &    72.8    &   77.7   &   58.7   &    73.2  &     95.5   &    54.1  &     60.9   &   106.0  &     85.9 \\
5731.76   &   4.26 &   -1.110 &    87.7    &   87.6   &   66.5   &    77.0  &    109.7   &    61.4  &     66.3   &   117.0  &     99.2 \\
\hline                  
\end{longtable}
} 

\onllongtab{13}{
\setlength\tabcolsep{2.5pt}
\begin{longtable}{lccccccccccc}

\caption{Line list and measured equivalent widths given in m$\rm \AA$}             
\label{t:ew6}   \\   
\hline\hline    
\noalign{\smallskip}
\noalign{\vskip 0.05cm} 
$\lambda$ [$\rm \AA$]& $\chi$ [eV]& log$gf$ & HD030608 & HD045415 & HD050778 & HD073017  & HD099648 & HD100920 & HD115478 & HD116976 & HD117220  \\
 (1)      & (2)    & (3)     & (4)      & (5)      & (6)      & (7)       & (8)      & (9)      & (10)     & (11)     &   (12)      \\		      
\hline 
\noalign{\vskip 0.05cm}
\endfirsthead

\multicolumn{12}{c}%
{{\tablename\ \thetable{} -- continued}} \\
\hline \hline 
\noalign{\smallskip}
\noalign{\vskip 0.05cm} 
$\lambda$ [$\rm \AA$]& $\chi$ [eV] & log$gf$ & HD030608 & HD045415 & HD050778 & HD073017  & HD099648 & HD100920 & HD115478 & HD116976 & HD117220  \\
 (1)      & (2)    & (3)     & (4)      & (5)      & (6)      & (7)       & (8)      & (9)      & (10)     & (11)     &   (12)      \\		      
\noalign{\vskip 0.05cm} 
\hline \\
\endhead
\endfoot
\hline
\endlastfoot
\noalign{\vskip 0.05cm}

{\bf FeI} &	   &	      & 	   &          &          &         &          &          &          &          &            \\ 
5560.21   &   4.43 &   -1.120 &     69.5   &   77.7   &   74.7   &   58.7  &	82.9  &    75.2  &    86.4  &	 98.4  &    36.9    \\
5633.95   &   4.99 &   -0.180 &     80.3   &   90.9   &   91.9   &   70.6  &	99.6  &    87.9  &   103.8  &	110.8  &    45.8    \\
5638.26   &   4.22 &   -0.800 &     99.6   &  108.9   &  113.0   &   85.8  &   118.9  &   102.3  &   121.5  &	122.2  &    61.3    \\
5651.47   &   4.47 &   -1.850 &     33.7   &   43.2   &   46.2   &   25.3  &	42.7  &    39.4  &    56.2  &	 56.5  &    11.4    \\
5662.52   &   4.18 &   -0.573 &    115.0   &  124.2   &  128.5   &   96.0  &   138.0  &   121.3  &   138.8  &	144.6  &    76.5    \\
5679.02   &   4.65 &   -0.810 &     69.7   &   81.1   &    ...   &   62.6  &	88.0  &    75.7  &    87.0  &	 96.4  &    42.4    \\
5731.76   &   4.26 &   -1.110 &     83.0   &   88.9   &   96.0   &   66.0  &   102.4  &    86.3  &   105.1  &	108.8  &    45.0    \\
\hline                  
\end{longtable}
} 

\onllongtab{14}{
\setlength\tabcolsep{2.5pt}
\begin{longtable}{lccccccccccc}

\caption{Line list and measured equivalent widths given in m$\rm \AA$}             
\label{t:ew7}   \\   
\hline\hline    
\noalign{\smallskip}
\noalign{\vskip 0.05cm} 
$\lambda$ [$\rm \AA$]& $\chi$ [eV] & log$gf$ & HD117818 & HD128188 & HD132345 & HD142948  & HD171496 & HD172223 & HD174116 & HD175219 & HD186378  \\
 (1)      & (2)    & (3)     & (4)      & (5)      & (6)      & (7)       & (8)      & (9)      & (10)     & (11)     &   (12)      \\		      
\hline 
\noalign{\vskip 0.05cm}
\endfirsthead

\multicolumn{12}{c}%
{{\tablename\ \thetable{} -- continued}} \\
\hline \hline 
\noalign{\smallskip}
\noalign{\vskip 0.05cm} 
$\lambda$ [$\rm \AA$]& $\chi$ [eV] & log$gf$ & HD117818 & HD128188 & HD132345 & HD142948  & HD171496 & HD172223 & HD174116 & HD175219 & HD186378  \\
 (1)      & (2)    & (3)     & (4)      & (5)      & (6)      & (7)       & (8)      & (9)      & (10)     & (11)     &   (12)      \\		      
\noalign{\vskip 0.05cm} 
\hline \\
\endhead
\endfoot
\hline
\endlastfoot
\noalign{\vskip 0.05cm}

{\bf FeI} &	   &	      &           &           &           &           &           &           &            &          &           \\     
5560.21   &   4.43 &   -1.120 &     69.4  & 	23.3  &      ...  & 	49.2  &     48.2  & 	99.1  &     77.1   & 	66.5  &     85.1   \\
5633.95   &   4.99 &   -0.180 &     81.0  & 	30.7  &    119.9  & 	57.4  &     61.4  & 	 ...  &     87.7   & 	84.2  &    105.0   \\
5638.26   &   4.22 &   -0.800 &     97.3  & 	49.1  &    141.0  & 	76.5  &     79.3  &    130.7  &    115.3   & 	94.6  &    123.4   \\
5651.47   &   4.47 &   -1.850 &     31.2  & 	 6.4  &     70.2  & 	15.0  &     16.1  & 	58.3  &     43.7   & 	30.0  &     53.6   \\
5662.52   &   4.18 &   -0.573 &    112.6  & 	64.9  &    161.9  & 	95.0  &     96.9  &    151.1  &    129.0   &   112.1  &    136.4   \\
5679.02   &   4.65 &   -0.810 &     75.0  & 	31.2  &    102.3  & 	55.6  &     52.9  & 	98.0  &     80.4   & 	68.4  &     90.2   \\
5731.76   &   4.26 &   -1.110 &     78.9  & 	29.2  &    121.8  & 	59.2  &     60.3  &    115.1  &     93.9   & 	77.7  &    100.7   \\
\hline                  
\end{longtable}
} 

\onllongtab{15}{
\setlength\tabcolsep{2.5pt}
\begin{longtable}{lcccccccc}

\caption{Line list and measured equivalent widths given in m$\rm \AA$}             
\label{t:ew8}   \\   
\hline\hline    
\noalign{\smallskip}
\noalign{\vskip 0.05cm} 
$\lambda$ [$\rm \AA$]& $\chi$ [eV] & log$gf$ & HD187195 & HD211075 & HD212320 & HD214376  & HD215030 &   HD221148  \\
 (1)      & (2)    & (3)     & (4)      & (5)      & (6)      & (7)       & (8)      & (9)          \\		      
\hline 
\noalign{\vskip 0.05cm}
\endfirsthead

\multicolumn{9}{c}%
{{\tablename\ \thetable{} -- continued}} \\
\hline \hline 
\noalign{\smallskip}
\noalign{\vskip 0.05cm} 
$\lambda$ [$\rm \AA$]& $\chi$ [eV] & log$gf$ & HD187195 & HD211075 & HD212320 & HD214376  & HD215030 &   HD221148  \\
 (1)      & (2)    & (3)     & (4)      & (5)      & (6)      & (7)       & (8)      & (9)          \\		      
\noalign{\vskip 0.05cm} 
\hline \\
\endhead
\endfoot
\hline
\endlastfoot
\noalign{\vskip 0.05cm}

{\bf FeI} &	   &	      &            &          &           &            &           &            \\ 	     
5560.21   &   4.43 &   -1.120 &     89.6   & 	74.0  &     70.0  & 	84.4   &    61.3   & 	82.9	\\
5633.95   &   4.99 &   -0.180 &    107.3   & 	85.4  &     88.2  & 	99.5   &    72.1   & 	90.9	\\
5638.26   &   4.22 &   -0.800 &    121.2   &   109.8  &    103.3  &    121.8   &    87.7   & 	 ...	\\
5651.47   &   4.47 &   -1.850 &     58.7   & 	40.0  &     32.0  & 	51.0   &    24.5   & 	51.7	\\
5662.52   &   4.18 &   -0.573 &    139.2   &   123.2  &    117.9  &    135.0   &    98.1   &   139.4	\\
5679.02   &   4.65 &   -0.810 &     87.9   & 	79.6  &     74.9  & 	89.8   &    62.4   & 	89.9	\\
5731.76   &   4.26 &   -1.110 &    103.5   & 	89.0  &     86.8  &    101.6   &    70.1   & 	95.7	\\
\hline                  
\end{longtable}
}

\end{document}